\documentclass[prb,aps,twocolumn]{revtex4}
\usepackage{amsfonts,amsmath,amssymb,latexsym}
\usepackage{epsfig}

\begin{document}

\title{ Full
counting statistics for noninteracting fermions:
Exact results and  
the Levitov-Lesovik formula}
\author {K. Sch\"onhammer}
\affiliation{Institut f\"ur Theoretische Physik, Universit\"at
  G\"ottingen, Friedrich-Hund-Platz 1, D-37077 G\"ottingen}

\date{\today}

\begin{abstract}
Exact numerical results for the full counting statistics (FCS) for a 
one-dimensional tight-binding model of noninteracting electrons
are presented without using an idealized measuring device. 
The two initially separate subsystems are connected at $t=0$ and 
the exact time evolution for the large but finite
combined system is obtained numerically
 via a finite dimensional determinant. Even for 
surprisingly short times the approximate description of the time
evolution with the help of scattering states agrees well with the
exact result for the local current matrix elements
and the FCS. An additional
approximation has to be made  to recover the Levitov-Lesovik
formula in the limit where the system size  becomes infinite and
afterwards the long time limit is addressed. The new derivation
of the Levitov-Lesovik formula is generalized to more general
geometries like a Y-junction enclosing a magnetic flux.

\end{abstract}

\maketitle

\section{Introduction}

The theory of noise in quantum transport in mesoscopic
systems is a very active field of research \cite{BB,Naza}.
In addition to the first few moments of the transmitted charge 
the full probability distribution can be studied, called
{\it full counting statistics} (FCS).
Neglecting the electron-electron interaction Levitov and Lesovik
presented an analytical result in the long time limit \cite{LL1,LL2}
using the cumulant generating function.
It was later
questioned pointing out that different measurement procedures may lead to
different answers for the distribution function\cite{LC}.
 Recently also results 
for models including the e-e-interaction locally were obtained.
\cite{GK,BKF}.

For the case of perfect transmission the Levitov-Lesovik formula
 \cite{LL1,LL2} yields at zero temperature
 a delta function for the probability
distribution $w(t,Q)$ of the transmitted charge with the position
increasing linearly with time. This is in obvious contradiction to the
fact that $w$ should only have weight at integer values.
In order to obtain the correct answer also in this limit it is
desirable to start from a formally exact expression. Such a
starting point was presented by Klich \cite{Klich} with the
help of a formula to calculate the trace over Fock space
for products of exponentials of one-particle operators.

Klich's formula can be viewed as the general expression for the
counting statistics of noninteracting fermions, at any given time,
and without any approximation\cite{Klich}. Nevertheless Klich's
attempt to derive the Levitov-Lesovik formula from his Eq. (13) is
not really satisfying. He performes the infinite time limit
to introduce the scattering matrix and notices that therefore
the information about the length of the time interval 
has to be put in by hand\cite{Klich}. As an alternative 
Muzykantskii and Adamov \cite{MA} start from Klich's formula and use
results from the theory of singular integral equations to proceed. 
They restrict themselves to zero temperature and switch on the bias 
adiabatically.
For the special case of an energy independent transmission probability
they recover the $T=0$ version of the Levitov-Lesovik formula
 from the solution of a matrix Riemann-Hilbert problem.
 From the
approximate solution of a more difficult  Riemann-Hilbert problem
they also obtain the subleading correction for their special
case. An extension of this approach to finite
temperatures was presented by Braunecker.\cite{BBr}  

The present paper takes a new look at the problem 
in several ways.
We first describe the ``experimental'' setup and its formal 
theoretical description at zero and finite temperatures.
For noninteracting fermions the central object is the time
dependent projection operator $P_R(t)$ onto one of the subsystems
 in the Hilbert space of one particle.
 By numerically calculating  $P_R(t)$  in section III
we present exact results for $w(t,Q)$ at $T=0$ for a 
 one-dimensional
tight-binding model. For perfect transmission
 the distribution has a nonzero width increasing
logarithmically with time. In section IV we
introduce scattering states to describe the long time limit.
By using the current operator at the connection of the subsytems
we obtain an excellent approximation for  $P_R(t)$.
The new derivation 
of the Levitov-Lesovik formula from Klich's exact formal expression
for arbitrary temperatures presented in section V 
 clearly shows the approximations involved.
We  also generalize the derivation to more
complex geometries like a Y-junction which encloses
a magnetic flux.  In order to make the paper selfcontained
we present an alternative proof of Klich's formula
using Wick's theorem in Appendix A.

\section{Counting statistics for noninteracting electrons}

\subsection{Zero temperature limit}

\noindent In the following we consider a system which consists
of two initially separate subsystems described by the Hamiltonians
$H_{0,a}$ with $a=L,R$. The labels stand for the ``left'' and
``right'' one-dimensional subsystems treated later. The number of
electrons in the subsystems in the initial state are $N_{0,a}$
and the total number is given by $N_{\rm tot}=N_{0,L}+N_{0,R}$.
We assume the intial state $|\Phi(0)\rangle$ to be an eigenstate
of the  $H_{0,a}$
\begin{equation}
\label{Phi0}
|\Phi(0)\rangle= |E_m^{N_{0,L}}\rangle \otimes |E_n^{N_{0,R}}\rangle~.
\end{equation} 
The time evolution for times greater than zero is described by
the Hamiltonian 
\begin{equation}
\label{Hamiltonian}
H=H_{0,L}+H_{0,R}+V_{LR}\equiv H_0 +V_{LR}~.
 \end{equation} 
The term $V_{LR}$ which couples the two subsystems will be specified 
later.
The probability distribution that $Q$ electrons are transfered to
the right system after time $t$ is given by
\begin{eqnarray}
\label{w1}
w_R(t,Q)&=& \langle \Phi(t)|\delta[Q-(
{\cal N}_R-N_{0,R})]|\Phi(t)\rangle \nonumber \\
&=& 
\langle \Phi(0)|
\delta[Q-({\cal N}_R(t)-N_{0,R})]|\Phi(0)\rangle .
\end{eqnarray}
Here ${\cal N}_R$ ist the particle number operator of the right system
and ${\cal N}_R(t) $ is the corresponding operator in the Heisenberg
picture. This probability distribution can be measured (conceptually simple by
``weighing'') when the connection between the subsystems is cut again
at time $t$.

 The integral represention of the delta function yields a 
form more convenient for the actual calculation  
 \begin{eqnarray}
\label{w2}
w_R(t,Q)=\frac{1}{2\pi}\int d\lambda e^{-i\lambda Q} g_R(t,\lambda)  ~,
\end{eqnarray}
where the information about the FCS is encoded in the
characteristic function
\begin{equation}
\label{gRT0}
g_R(t,\lambda)=
 \langle \Phi(0)|   e^{i\lambda {\cal N}_R(t) } | \Phi(0) \rangle
  e^{-i\lambda N_{0,R}}~.
\end{equation}
This result is valid also for interacting electrons.

 The evaluation
simplifies considerably for noninteracting electrons as the
expectation value in Eq. (\ref{gRT0}) can be expressed as a 
$N_{\rm tot}\times N_{\rm tot} $ determinant in the space of the
initially occupied one-particle states. \cite{CN} As ${\cal N}_R(t)  $ is 
a one-particle operator $e^{i\lambda{\cal N}_R(t)} $
   acts on the one-particle states occupied
in the initial Slater determinant as $e^{i\lambda P_R(t)}$ where
$P_R(t)$ is the time dependent projection operator
\begin{equation}
P_R(t)=e^{iht}P_Re^{-iht}=P_R^2(t)
\end{equation}
with  $P_R$ 
the projection operator 
 in the Hilbert space of one
particle onto the states in the right reservoir
\begin{equation}
P_R=\sum_j |k_j^R\rangle \langle k_j^R| = \sum_m |m,R\rangle \langle m,R|
\end{equation}
and $h$ ist the full Hamiltonian in the Hilbert space of a single
particle.
The $|k_j^R\rangle $ describe standing waves in the right
subsystem and the $m$ in  $|m,R\rangle $ denote the site indices. 

  If we introduce  
the projection operator $\bar n_0$  
\begin{eqnarray}
\bar n_0
=\sum_{k,a}|k^a\rangle f_{k,a}\langle k^a|\equiv \bar n_{0,L}
+\bar n_{0,R} \nonumber~,
\end{eqnarray}
where the $f_{k,a}$ are the ($T=0$) Fermi functions of the
subsystems in the initial state the operator
\begin{equation}
\bar P_R(t) \equiv  \bar n_0  P_R(t)\bar n_0~.
\end{equation} 
determines the zero temperature FCS. 
Using  $e^{i\lambda P_R(t)}=\hat 1 +(e^{i\lambda}-1) P_R(t) $
we obtain
\begin{eqnarray}
\label{KlichT0}
g_R(t,\lambda)
&=& e^{-i\lambda N_{0,R}}
  {\rm det}_{\bar n_0}\left [\bar n_0 +(e^{i\lambda}-1) \bar P_R(t)
  \right ]~,
 \end{eqnarray}
where $ {\rm det}_{\bar n_0} $ is a determinant in the subspace
of the initially occupied one particle states. 
If we denote the  eigenvalues of $\bar P_R(t) $
 by $p_m(t) $ the
characteristic function takes the form
\begin{eqnarray}
\label{Klich01}
g_R(t,\lambda)
&=&e^{-i\lambda N_{0,R}}\prod_{m=1}^{N_{\rm tot}}
 \left [1+(e^{i\lambda}-1)p_m(t)\right ] \nonumber \\
&=&\sum_{n=0}^{N_{\rm tot}}c_n(t)e^{i(n-N_{0,R})\lambda}~.
\end{eqnarray}
The coefficients $c_n(t)$ can be calculated recursively from the product
representation.  
This leads to the probability distribution
\begin{equation}
w_R(t,Q)= \sum_{n=0}^{N_{\rm tot}}c_n(t)\delta\left
  (Q-(n-N_{0,R})\right )~.
\end{equation}
To obtain exact results for the FCS one first has to calculate $\bar
P_R(t)$ and then obtain its eigenvalues $p_m(t)$.

\subsection{Finite temperatures}

The starting point for finite temperatures is again given by
Eq. (\ref{gRT0}).
We take the factor $e^{-i\lambda N_{0,R}}$ inside the expectation
value and replace it by $  e^{-i{\cal N}_R(t)\lambda }  $.  
We then perform a grand canonical averaging corresponding to the 
statistical operator
\begin{eqnarray}
\rho_0&=& \frac{e^{-\beta_L (H_{0,L}-\mu_L{\cal N}_L)}}
{ {\rm Tr} e^{-\beta_L(H_{0,L}-\mu_L{\cal N}_L)} }\otimes 
\frac{e^{-\beta_R(H_{0,R}-\mu_R{\cal N}_R)}}
{{\rm Tr} e^{-\beta_R(H_{0,R}-\mu_R{\cal N}_R)}} \nonumber \\
&\equiv& e^{-\bar H_0}/\bar Z_0
 \end{eqnarray}
The probability distribution $w$ is then given by 
 \begin{eqnarray}
\label{w3}
w_R^{gc}(t,Q)=\frac{1}{2\pi}\int d\lambda e^{-i\lambda Q} 
\left \langle e^{i\lambda {\cal N}_R(t)}
  e^{-i\lambda {{\cal N}_{R}}}\right \rangle~,
\end{eqnarray}
where $\langle \cdot \rangle$ denotes the averaging with
the statistical operator $\rho_0$.
The information about the FCS is therefore encoded in the
characteristic function
\begin{equation}
\label{gR}
g_R(t,\lambda)=
\left \langle e^{i\lambda {\cal N}_R(t) }  e^{-i\lambda {\cal N}_{R}}
 \right \rangle~.
\end{equation}
Again this result is valid also for interacting electrons. 

For noninteracting fermions the characteristic function can again
be expressed as a determinant but now in the full one particle
Hilbert space. One can either use Klich's trace formula \cite{Klich}
(Eq. (\ref{Klichf})) or use the proof using Wick's theorem presented
in apppendix A. This yields 
\begin{eqnarray}
\label{Kli}
g_R(t,\lambda)= {\rm det}\left [\hat 1 +\left (e^{i\lambda P_R(t)}
e^{-i\lambda P_R}  -\hat 1\right )\bar n_0
  \right  ] ~,
\end{eqnarray}
where the Fermi functions in the definition of $\bar n_0$ 
are now the ones for finite temperatures. This implies that
 $\bar n_0$ is no longer a projection operator.
At $T=0$ Eq. (\ref{Kli}) reduces to  Eq. (\ref{KlichT0}).

As the next step we rewrite Eq. (\ref{Kli}) in the form 
\begin{eqnarray} g_R(t,\lambda)
={\rm det}\left [\hat 1 -\bar n_0+e^{i\lambda P_R(t)}\left (\bar n_{0,L}
+e^{-i\lambda }\bar n_{0,R}\right )\right ] \nonumber
\end{eqnarray}
In contrast to statements in the literature 
\cite{LL2} it is useful to write 
 $P_R(t)$ as a sum 
\begin{eqnarray}
\label{deltaP} 
P_R(t)&=&P_R(0)+\int_0^t\dot P_R(t') dt'\equiv P_R+\delta P_R(t)~.~~
\end{eqnarray}
With $  e^{i\lambda
  P_R(t)}
=\hat 1+(e^{i\lambda}-1)(P_R+\delta P_R(t))  $
we obtain 
\begin{eqnarray} 
e^{i\lambda P_R(t)} \!\!\!&(&\!\!\!\bar n_{0,L}
+  e^{-i\lambda }\bar n_{0,R} )- \bar n_0 \\ 
&=& \delta P_R(t)\left [ \bar n_{0,L}\left ( e^{i\lambda}-1 \right )
-\bar n_{0,R}\left ( e^{-i\lambda}-1 \right ) \right ] \nonumber \\
&\equiv& \delta P_R(t)\left [\bar a_L(\lambda)-\bar a_R(\lambda) \right ]~. \nonumber
\end{eqnarray}
The general result for the characteristic function then
takes the form
\begin{eqnarray}
\label{KlichLL} 
g_R(t,\lambda)={\rm det}\left [\hat 1 +\delta P_R(t)
\left (\bar a_L(\lambda)-\bar a_R(\lambda) \right )\right ]~.
\end{eqnarray}
As shown in section VI this is the starting point for a 
transparent derivation of the
approximate (long time limit) Levitov-Lesovik formula
\begin{eqnarray}
\label{LLformula}
\ln{g_R^{(L)}}=\frac{t}{2\pi}\int  \!\!\!&d\epsilon& \!\!\! 
  \ln  \{1+T(\epsilon) [
(e^{i\lambda}-1)f_L(\epsilon)(1-f_R(\epsilon)) \nonumber \\
 \!\!\!&+& \!\!\!  (e^{-i\lambda}-1)f_R(\epsilon)(1-f_L(\epsilon)) ] \},~
\end{eqnarray}
where $T(\epsilon)$ is the transmission probability for
the scattering of a single particle between the two subsystems.

Note that compared to the zero temperature result the operator added to the 
unit operator  in the determinant
in Eq. (\ref{KlichLL}) cannot be written as a product of a function
of $\lambda$ and an operator independent of $\lambda$. Therefore the
numerical effort for an exact numerical calculation of $w_R(t,Q)$ is
much higher than at $T=0$.

\section {Exact numerical results for $T=0$}

\subsection{The model}

In this section we present exact numerical results for the
probability distribution $w(t,Q)$ for a one dimensional tight
binding model. The  generalization for the model with more than two
leads discussed in section VI is straightforward.

 The unperturbed one particle Hamiltonians of the subsystems
 are given by
\begin{eqnarray}
\label{hamiltonian}
h_{0,L}&=&-\sum_{m=-(M_L-1)}^{-1}\tau_m (|m\rangle\langle m+1|+H.c.)
+v_0|0\rangle \langle 0|\nonumber \\
h_{0,R}&=&-\frac{B}{4}\sum_{m=1}^{M_R-1} (|m\rangle\langle m+1|+H.c.)~.
\end{eqnarray}
The number of sites $m$ in the subsystems are given by $M_a$.
In the explicit calculations we later specialize to $\tau_m=B/4$ for
$m=-(M_L-1),...,-2$ and $\tau_{-1}=t_L$.\\  

 The one particle eigenstates $|k_j^R\rangle$ of the right subsystem
are standing waves
\begin{eqnarray}
\label{standing1}
\langle m|k_j^R\rangle=\sqrt{\frac{2}{M_R+1}}\sin(k_j^Rm)~, 
~~~  k_j^R=\frac{j\pi}{M_R+1}
\end{eqnarray}
with the integer $j\in[1,M_R]$. The corresponding energies are
given by $\epsilon_{k_j^R}=-(B/2)\cos k_j^R$, i.e. the total band
width is given by $B$.

In order to obtain a stationary current flow in  
 the long time limit $t \to \infty$ one first has to
take 
the limit $M_a\to \infty$.
 In this limit the quantum numbers $k_j$ become continuous with
$k\in [0,\pi]$ and one has
 to go over to the delta function
normalization of the states
\begin{equation}
\label{standing2}
\sqrt{\frac{M_R+1}{\pi}}|k_j^R\rangle \to |k,R\rangle
\end{equation}
with $\langle k,R|k',R\rangle =\delta(k-k')$.

 In the left subsystem the eigenstates $|k_i^L\rangle$ are pure
sine-waves only in the special case $v_0=0$ and $\tau_m=const.$

 The coupling between the two subsystems is described by the
hopping term
\begin{equation}
\label{connection}
v_{LR}=-t_R \left (|0\rangle\langle 1|+H.c.\right)~.
\end{equation}
The special case $v_0=0$ and $\tau_m=const.$ corresponds to two 
subsystems which, apart from their initial filling, are identical.
The numerical results presented in this section are for this 
simplest case. The model with $\tau_{-1}=t_L$ and $v_0$ taking arbitrary
values is a model which allows resonant scattering through 
the  ``quantum dot'' at site zero. Some properties of this model 
simplify in the ``wide band limit'' $B\to \infty$.

\subsection{Results for finite times}

As discussed in the previous section one has to address the eigenvalue
problem of the operator $\bar P_R(t)$ in order to calculate the  
probability distribution $w(t,Q)$.
For the calculation of the eigenvalues $ p_m(t)$ of  $\bar P_R(t)$ we use
\begin{eqnarray}
\label{PR1}
\langle k_{\mu}^a|\bar P_R(t)|k_{\mu'}^{a'}\rangle =
\sum_{m=1}^{M_R} \langle k_{\mu}^a(t)|m\rangle
 \langle m|k_{\mu'}^{a'}(t)\rangle~,
 \end{eqnarray}
where $\mu,\mu' $ label the occupied one particle states.
 With the help of the spectral representation of the full
Hamiltonian $h$ 
\begin{equation}
h=\sum_i|k_i\rangle \epsilon_{k_i}\langle k_i|
\end {equation}
the time dependent quantities in Eq.(\ref{PR1}) can be 
expressed as sums
\begin{equation}
 \langle m|k_{\mu'}^{a'}(t)\rangle
=\sum_i \langle m|k_i\rangle e^{-i\epsilon_ {k_i}t}\langle k_i
|k_{\mu'}^{a'}\rangle~.
\end {equation}
In order to obtain the eigenvalues the resulting 
 $N_{\rm tot} \times N_{\rm tot}  $ matrix has to be diagonalized
numerically.\\

As discussed in more detail in the next section the results are almost
independent of the size of the system and only depend on the values of
the chemical potential $\mu_L$ and $\mu_R$ for times smaller than 
the time it takes  the
``charge fronts'' which move into the subsystems after connecting
them to return to the  the connection point after having been reflected
at the boundaries of the system. \\

 We always
consider the case $N_{0,L}\ge N_{0,R} $, i.e. the net particle flow 
is from left to right. As 
 $ P_R(t)$ is a projection operator the 
eigenvalues  $p_m(t)$ of $\bar P_R(t)$ obey the 
bounds $0\le p_m (t) \le 1$.
 In the figures we show the eigenvalues
$p_m(t)$ in
descending order.

\vspace{0.5cm}    
\begin{figure}[tb]
\begin{center}
\vspace{-0.0cm}
\leavevmode
\epsfxsize7.5cm
\epsffile{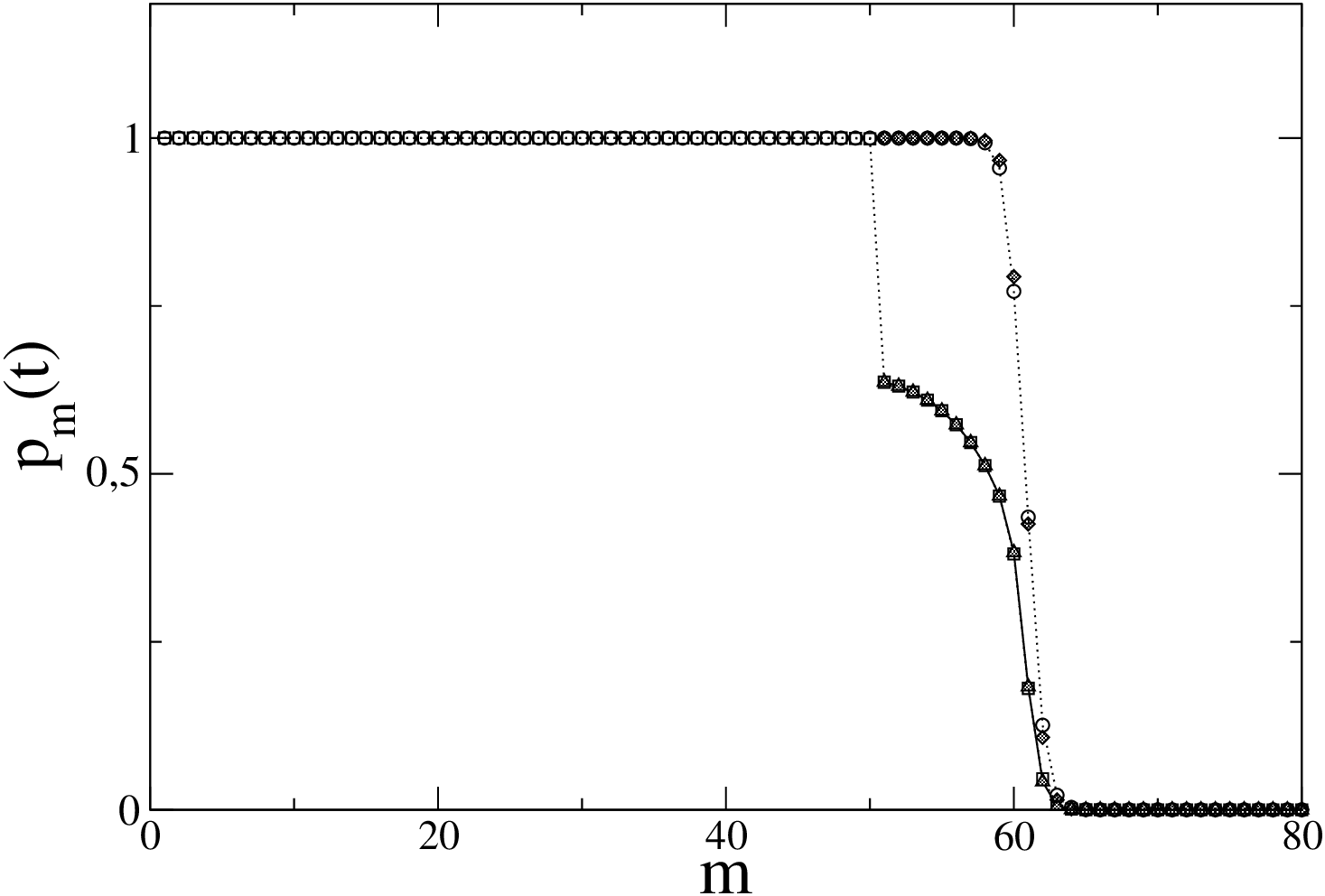}
\caption {Eigenvalues $p_m(t)$ of $\bar P_R(t)$ 
 for a system with  $M_L=M_R=100$, $N_{0,L}=80, N_{0,R}=50 $ and $\hat t
\equiv tB/4=40$ for $\tilde t_R \equiv 4t_R/B=1.0$ (circles) and 
 $\tilde t_R=0.5$ (squares).
 Starting with $m=51$ they are compared to the $N_{\rm
   diff}$
eigenvalues of $\bar P_{R,\rm diff}(t)$ defined in the text:
$\tilde t_R=1.0$
(diamonds) and $\tilde t_R=0.5$ (triangles up). }
\label{Levitov01}
\end{center}
\end{figure}

For $t=0$ the matrix $\bar P_R(0)$
is diagonal with $N_{0,R}$ eigenvalues $1$ and $N_{0,L}$ eigenvalues
$0$. For $t>0$ the  $N_{0,R}$ eigenvalues almost identical to $1$
persist.
As shown in Fig. \ref{Levitov01} for perfect transmission
$N_t$ additional eigenvalues almost identical to
$1$ occur where the value $N_t \sim t$ is
 discussed in detail later.\cite{Riecke} 
The crossover from eigenvalues close to one to those close to zero
is smooth. 
For non-perfect transmission the eigenvalues which significantly differ
from $1$ and $0$ start at $N_{R,0}+1$ and approximately extend to 
$N_{R,0}+N_t$. 

If one has a look at the Levitov-Lesovik formula Eq. (\ref{LLformula})
 for $T=0$
\begin{eqnarray}
\label{LeLezero}
\ln{g_R^{(L)}(t,\lambda)}=\frac{t}{2\pi}\int_{\mu_R}^{\mu_L}d\epsilon
 \ln  \left [1+T(\epsilon) (e^{i\lambda}-1)\right ]~.
\end{eqnarray}
one expects the states below $\mu_R$ which are initially occupied in 
both subsystems to be irrelevant. We therefore also calculated 
 the eigenvalues of the 
$N_{\rm diff}\times N_{\rm diff} $ submatrix $\bar P_{R,\rm diff}(t)$
in the space of the $N_{\rm diff}=N_{0,L}-
N_{0,R} $ states of the left system
in the energy window $[\mu_R,\mu_L]$.
As can be seen in Fig. \ref{Levitov01} these eigenvalues
deviate little from the eigenvalues of the full $N_{\rm tot}\times
N_{\rm tot}$ matrix.

\begin{figure}[tb]
\begin{center}
\leavevmode
\epsfxsize7.5cm
\epsffile{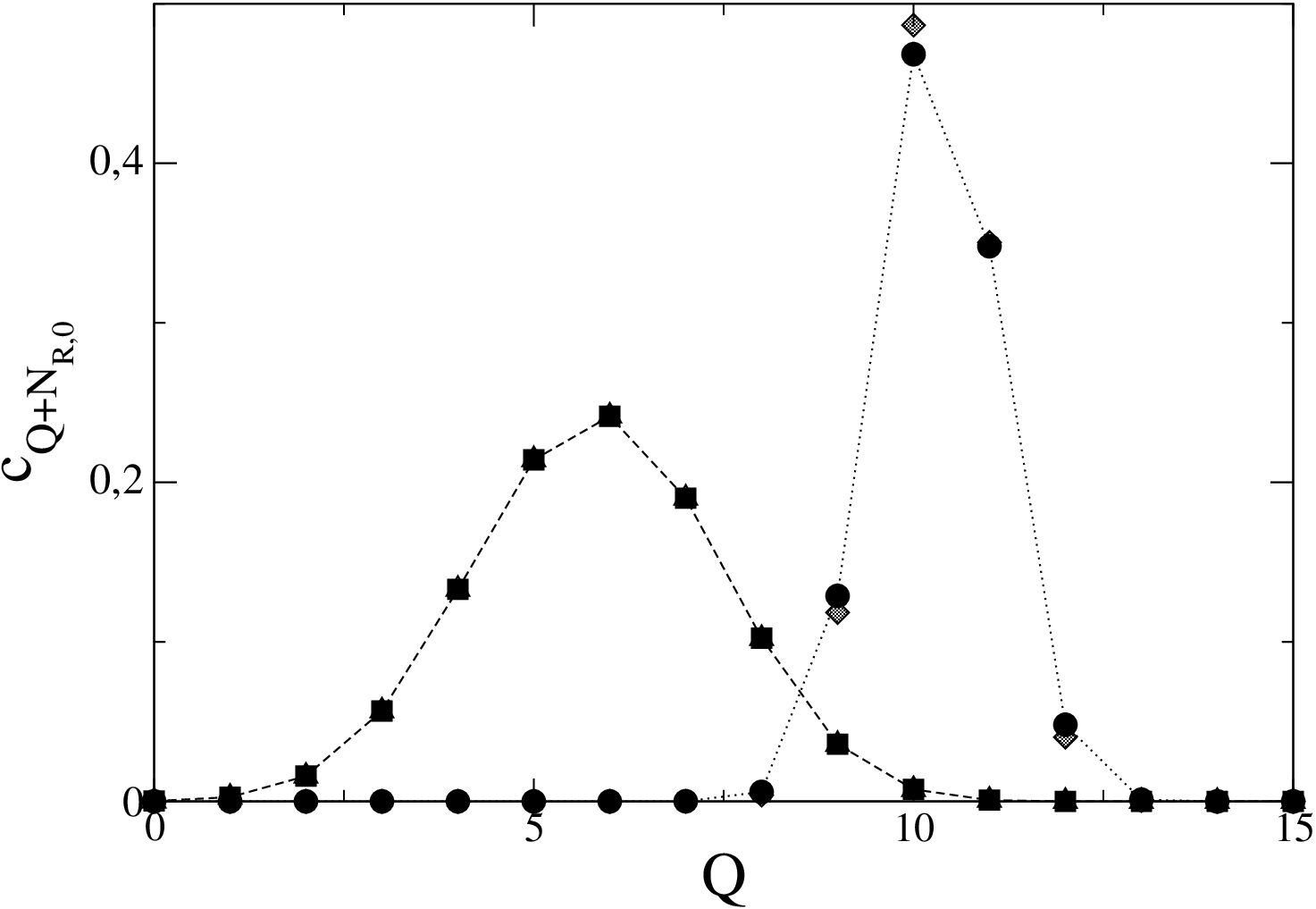}
\caption {Weights of the probability ditribution $w_R(t,Q)$ for the 
two parameter sets of Fig. \ref{Levitov01}. From the eigenvalues of
$\bar P_R(t)$: $\tilde t_R=0.5$ (filled squares), $\tilde t_R=1.0$ (filled
circles)
and the eigenvalues of  $\bar P_{R,\rm diff}(t)$:
  $\tilde t_R=0.5$ (triangles up), $\tilde t_R=1.0$ (diamonds).}
\label{Levitov02}
\end{center}
\vspace{0.0cm}
\end{figure}

The weights of the probability distribution corresponding to the
parameters
of Fig. \ref{Levitov01} are shown in Fig. \ref{Levitov02} .
 For the non-ideal transmission
($\tilde t_R\equiv 4t_R/B=0.5$) the distribution calculated from the ``extra
electrons''
only, agrees very well with the one from the full calculation.
For the perfect transmission case slight deviations show up.
  
 If one doubles the size of the system and
the number of electrons but leaves $t$ fixed, the result for
$w_R(t,Q)$
is almost identical to the one shown in Fig. \ref{Levitov02}.
Our approach using finite systems and nevertheless see 
``asymtotic'' properties in a time window is very useful also
for interacting fermions for which effective methods have been
worked out recently to describe the time dependence for finite 
systems. Promising techniques are the time dependent density
matrix renormalization group (DMRG) \cite{DMRG,PSE} and the 
time dependent numerical renormalization group (NRG).\cite{NRG}

 \vspace{1.0cm}    

Note that the $T=0$ Levitov-Lesovik formula predicts zero width
of the probability distribution for the perfect transmission case
which is in contrast to the exact numerical result in
Fig. \ref{Levitov02}.
 To study
the non-zero
width as a function of time one can calculate the second order cumulant.

\subsection{Low order cumulants}
 The low order cumulants $\kappa_i$ of $w_R(t,Q)$ 
are obtained from the Taylor expansion of $\ln g_R(t,Q)$. 
Using ${\rm det}~a=\exp{({\rm tr}\ln a)}$ in Eq. (\ref{KlichT0}) one
obtains for $i$ up to $4$

\begin{eqnarray}
\label{cumulantszero}
\kappa_1&=& {\rm tr}\bar P_R -N_R^0 \\
\kappa_2&=& {\rm tr}\bar P_R- {\rm tr}\bar P_R^2 \nonumber \\
\kappa_3&=& {\rm tr}\bar P_R- 3{\rm tr}\bar P_R^2
            +2{\rm tr}\bar P_R^3 \nonumber \\
 \kappa_4&=& {\rm tr}\bar P_R- 7{\rm tr}\bar P_R^2
           +12{\rm tr}\bar P_R^3   -6 {\rm tr}\bar P_R^4 \nonumber ~,
\end{eqnarray}
In the long time limit discussed in the next section analytical
results for the low order cumulants can be obtained.

\begin{figure}[tb]
\begin{center}
\vspace{-0.0cm}
\leavevmode
\epsfxsize7.5cm
\epsffile{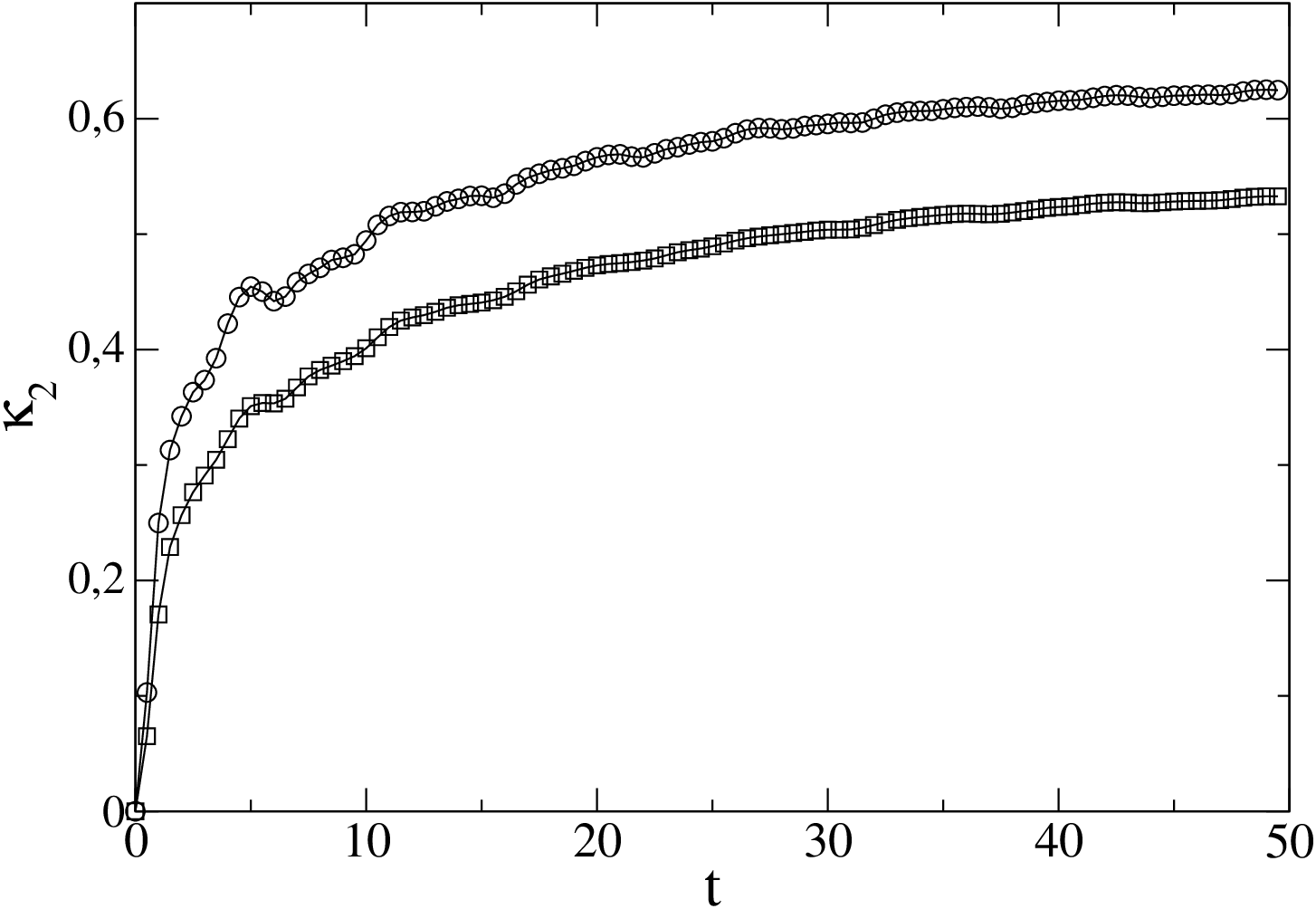}
\caption {Second order cumulant $\kappa_2$
for perfect transmission at $T=0$ as a function of time for
  the system parameters of Figs. \ref{Levitov01} and \ref{Levitov02}.
The circles show the result using the matrix $\bar P_R(t)$, while the
squares are the results for the   matrix $\bar P_{R,\rm diff}(t)$
which only treats the electrons in the energy window $[\mu_R,\mu_L]$.}
\label{Levitov03}
\end{center}
\end{figure}

 In Fig. \ref{Levitov03}
we show results for $\kappa_2$ for 
perfect transmission and the parameters
of Fig. \ref{Levitov01}   . The circles correspond
 to the full calculation and the 
squares to the result using $\bar P_{R,\rm{diff}}(t)$. It will be
shown analytically
in the next section that   $\kappa_2$ increases logarithmically
at $T=0$ in the long time limit for perfect transmission.
The numerical results for $\kappa_i$ with $i\ge 3$ show no logarithmic
increase. Apart from small transient effects these cumulants stay
close to zero.

The Levitov-Lesovik formula for $T=0$ predicts zero width
of the probability distribution also for the zero bias case
$\mu_L=\mu_R$ for arbitrary transmission. Absolutely no charge
transport is predicted in contrast to the exact numerical results
not shown here.

\vspace{0.5cm}    

\section{The long time limit}

\subsection{Introduction of scattering states}

We now examine if 
  $ P_{R}(t)$  can be
discussed analytically  in the long time limit.
For the calculation of the matrix elements of $ P_R(t)$ 
it is useful to split off a  phase factor
in the time dependent one particle states 
 and define
\begin{equation}
|k_j^a\rangle_t\equiv e^{i\epsilon_{k_j}^at}|k_{j}^a(t)\rangle=
e^{-i(h-\epsilon_{k_j}^a)t}|k_{j}^a\rangle
\end{equation}
It is known from scattering theory \cite{Taylor} that in the long time limit 
(after taking the limit $M_a \to \infty$) the states $|k,a\rangle_t$
converge to the scattering states.  

\noindent  This can be seen using an Abelian
limit procedure
\begin{eqnarray}
\label{scatstate}
\lim_{t \to \infty} e^{-i(h-\epsilon_{k,a})t}|k,a\rangle
 &=&
\lim_{\eta \to 0}\eta \int_0^\infty e^{-\eta t }
 e^{-i(h-\epsilon_{k,a})t}|k,a \rangle dt \nonumber \\
 &=& \lim_{\eta \to 0} \frac{i\eta}{\epsilon_{k,a}-h+i\eta}|k,a
 \rangle \nonumber \\
&=& |k,a+\rangle
\end{eqnarray}
The state on the rhs of the last equality is just the scattering
state with outgoing scattered waves.

The explicit form of the scattering states for our Hamiltonian $h$
with $\tau_m=B/4$ for $m\le -1$ is given by
\begin{eqnarray}
\label{StreuL}
\langle m|k,L+\rangle
&=&\frac{1}{\sqrt{2\pi}}(e^{ikm}+r_ke^{-ikm})~,~~~m\le -1 \nonumber \\
&=&\frac{1}{\sqrt{2\pi}}t_ke^{ikm}~,~~~m\ge 1
\end{eqnarray} 
with $r_k$ and $t_k$ the reflection and transmission amplitude.
Similarly the scattering states with incoming wave from the right
are given by
\begin{eqnarray}
\label{StreuR}
\langle m|k,R+\rangle
&=&\frac{1}{\sqrt{2\pi}}(e^{-ikm}+\tilde r_ke^{ikm})~,~~~m\ge 1 \nonumber \\
&=&\frac{1}{\sqrt{2\pi}}\tilde t_ke^{-ikm}~,~~~m\le -1~.
\end{eqnarray} 

\noindent If one returns to the description of the finite system at finite
times the question arises how well the approximation
\begin{equation}
\label{appr}
\langle m|k_j^a(t)\rangle\approx \sqrt{\frac {\pi} {N_a+1}}
\langle m|k_j,a+\rangle e^{-i\epsilon_{k_j}^at}
\end{equation}
works.

\vspace{0.8cm}    
\begin{figure}[hbt]
\begin{center}
\vspace{-0.0cm}
\leavevmode
\epsfxsize7.5cm
\epsffile{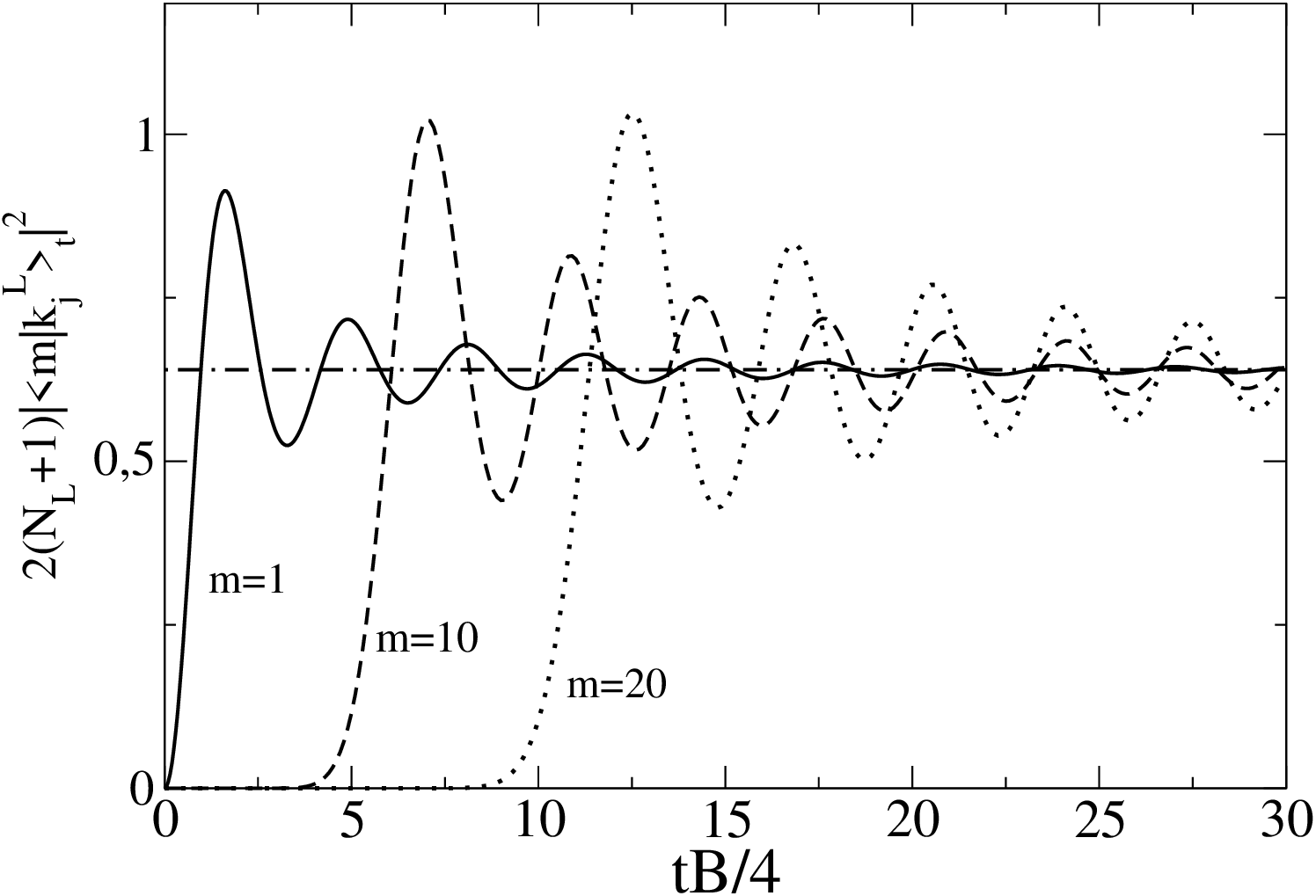}
\caption {This figure shows how the long time
 approximation in Eq.(\ref{appr}) works for finite times. The quantity
plotted should converge to the value $|t_{k_j}|^2$ for all values
$m>0$ of the site index. Shown is the result for a state in the
middle of the band for a system with  $M_L=M_R=100$
and $\tilde t_R \equiv 4t_R/B=0.5$. For the times shown
the results agree within the plotting accuracy with corresponding
calculations for   $M_L=M_R\gg 100$. }
\label{Levitov04}
\end{center}
\vspace{0.0cm}
\end{figure}

 In Fig. \ref{Levitov04} we show $2(N_L+1)|\langle
m|k_j^L\rangle_t|^2$  for different values 
$m>0$ of the site index for identical subsystems with $M_L=M_R=100$, 
hopping matrix element $\tilde t_R \equiv 4t_R/B=0.5$
 and a state in the middle of the
band. In the approximation of Eq. (\ref{appr}) this quantity should
be time independent and given by the transmission probability $|t_{k_j}|^2$ . 
The figure shows that the larger the value of $m$ the later the oscillatory convergence
sets in.

 It would be totally
wrong to use Eq. (\ref{appr}) in Eq. (\ref{PR1}) for all sites of
the subsystems. The only time dependence would then result 
from the factors
$e^{i\epsilon_{k_{\mu,a}}t}e^{-i\epsilon_{k_{\mu',a'}}t}$.
They could be gauged away by absorbing them in the definition of
the basis states making the eigenvalues $p_m(t)$ time independent.
This would imply that $g_R(t,\lambda)$ and
with it the FCS is time independent.  
This shows clearly that one should apply Eq. (\ref{appr}) only
to the calculation of observables localized in the neighborhood
of the junction.\\

 It is therefore
useful to express $ P_R(t)$ as an integral over the current 
operator between the two subsystems as in Eq. (\ref{deltaP})
\begin{eqnarray}
 P_R(t) =P_R(0)+\int_0^t j_{0\to 1}(t') dt'\equiv P_R+\delta P_R(t)
\end{eqnarray} 
with the current operator given by 
\begin{eqnarray}
 j_{0\to 1}=it_R\left (|1\rangle \langle 0|-|0\rangle \langle 1|\right )~.
\end{eqnarray}
The matrix elements of $ P_R(t)$ are therefore given by
\begin{eqnarray}
\label{current1}
\langle k_j^a| P_R(t)|k_{j'}^{a'}\rangle&=&
\delta_{a,a'}\delta_{a,R}\delta_{j,j'} \nonumber \\
&~+&\int_0^t\langle k_{j}^a(t')|
 j_{0\to 1}   |k_{j'}^{a'}(t')\rangle dt'~.
\end{eqnarray} 

As expected from Fig. \ref{Levitov04} the use of the approximation in
Eq. (\ref{appr}) to evaluate the matrix elements of the local
observable $j_{0\to 1} $ works surprisingly well even for rather
short times. With this replacement the time integration in
Eq. (\ref{current1}) can be carried out and one obtains
the very useful approximation also for finite systems
\begin{eqnarray}
\label{appr2}
\langle k_{j}^a| P_R(t)|k_{j'}^{a'}\rangle\approx
\delta_{a,a'}\delta_{a,R}\delta_{j,j'}~~~~~~~~~~~~~~~~~~~~~~~~~  \\
+ \pi \frac{\langle k_j,a+|j_{0\to 1}|k_{j'},a'+\rangle }
{\sqrt{(M_a+1)(M_{a'}+1)}}\frac{e^{i(\epsilon_{k_{j,a }}
-\epsilon_{k_{j'},a' })t}-1}{i(\epsilon_{k_{j,a }}
-\epsilon_{k_{j'},a' }) }.\nonumber
\end{eqnarray}

For the calculation of the current matrix elements
we also have to know how $\langle 0|k,a+\rangle$ is related
to the scattering amplitudes in Eqs. (\ref{StreuL})
and  (\ref{StreuR}). The Schr\"odinger equation $\langle
1|h|k,a+\rangle=\epsilon_{k,a}\langle 1|k,a+\rangle$ yields
\begin{eqnarray}
\label{zerosite}
t_R\langle 0|k,L+\rangle=\frac{t_k}{\sqrt{2\pi}}\frac{B}{4},~~
t_R\langle 0|k,R+\rangle=\frac{1+\tilde r_k}{\sqrt{2\pi}}\frac{B}{4} .~
\end{eqnarray} 
This allows to express the current matrix elements in Eq.(\ref{appr2})
in terms of the $t_k$ and $\tilde r_k$, e.g.
\begin{eqnarray}
\label{current2}
\langle k,L+|j_{0\to 1}|k',L+\rangle=\frac{i}{2\pi}t_k^*t_{k'}   (e^{-ik}-e^{ik'})
\frac{B}{4}~.
\end{eqnarray}
and 
\begin{eqnarray}
\label{current3}
\langle k,L+|j_{0\to 1}|k',R+\rangle&=&\frac{i}{2\pi}[t_k^*\tilde r_{k'}
  (e^{-ik}-e^{ik'}) \nonumber \\
&+&t_k^* (e^{-ik}-e^{-ik'})]
\frac{B}{4}~.
\end{eqnarray}

These expressions simplify in the in the wide band limit
$|\mu_L-\mu_R|\ll B$ for the scattering states in the energy
window between $\mu_L$ and $\mu_R$. 
Then
one can replace $k$ and $k'$ by $\bar k_F\equiv (k_{F,L}+k_{F,R})/2$. This yields
  \begin{eqnarray}
\label{appr3}
\langle k,L+|j_{0\to 1}|k',L+\rangle&\approx&\frac{v(\bar
  k_F)}{2\pi}t_k^*t_{k'}\nonumber \\
\langle k,L+|j_{0\to 1}|k',R+\rangle&\approx&\frac{v(\bar
  k_F)}{2\pi}t_k^*\tilde r_{k'}
\end{eqnarray}
with the Fermi velocity $v_F\equiv v(\bar k_F)=2B\sin \bar k_F$. 
As the transmission amplitudes may have
a rapid variation in this energy window the replacement $k,k'\to \bar k_F$
has not been made for the factors involving the scattering
amplitudes.
 
\subsection{Low order cumulants}

\noindent As for the Hamiltonian in Eq.(\ref{hamiltonian})
 the transmission and 
reflection amplitudes can easily be calculated
 (see the appendix), the
matrix elements of $ P_R(t)$ in
the scattering approximation
 Eq. (\ref{appr2}), called  $ P^s_R(t)$ in the following,  are known
explicitely. Without further approximations the eigenvalues of
 this matrix have to be calculated
numerically. The low order cumulants $\kappa_i$ of $w_R(t,Q)$ on the other 
hand 
can be discussed analytically.
We begin with the $T=0$ limit where the cumulants have the simple form 
presented in Eq. (\ref{cumulantszero})

\subsubsection{Results for $T=0$}

 Using Eqs. (\ref{cumulantszero}) and  (\ref{appr2})   the explicit form
of the time derivative of the first order cumulant for $T=0$ is given by  
\begin{eqnarray}
\label{kappa1p}
\dot \kappa_1(t)&=&
\sum_{\mu,a}\langle k_{\mu}^a(t)|
 j_{0\to 1}   |k_{\mu}^{a}(t)\rangle  \nonumber \\
&\approx&
\sum_{\mu,a}\frac{\pi}{N_a+1}  \langle k_\mu,a+|j_{0\to 1}|k_{\mu},a+\rangle 
\end{eqnarray}
For symmetric subsystems ($\tau_m=B/4$,~$v_0=0$ and $M_L=M_R$)   the 
contributions of the states with energy smaller than $\mu_R(<\mu_L)$
which are occupied in both subsystems exactly cancel and only the
states $|k_\mu^L\rangle $ in the energy window $[\mu_R,\mu_L]$
contribute.
In the thermodynamic limit $
\pi/ (N_a+1)/ \sum_{k_a}(\cdot)\to
\int (\cdot)dk=\int (\cdot) d\epsilon/v(\epsilon)$
the cancellation of the contributions from the states below $\mu_R$
holds generally in the approximation of using the scattering states
 and one obtains using Eq. (\ref{current2})
\begin{eqnarray}
\label{LB}
\dot \kappa_1(t)\approx \frac{1}{2\pi}\int_{\mu_R}^{\mu_L}T(\epsilon)d\epsilon~.
\end{eqnarray}
This is the zero temperature version of the Landauer-B\"uttiker formula
with $T(\epsilon)=|t_{k(\epsilon)}|^2$ the transmission probability.\cite{LaBu}

\begin{figure}[hbt]
\begin{center}
\vspace{-0.0cm}
\leavevmode
\epsfxsize7.5cm
\epsffile{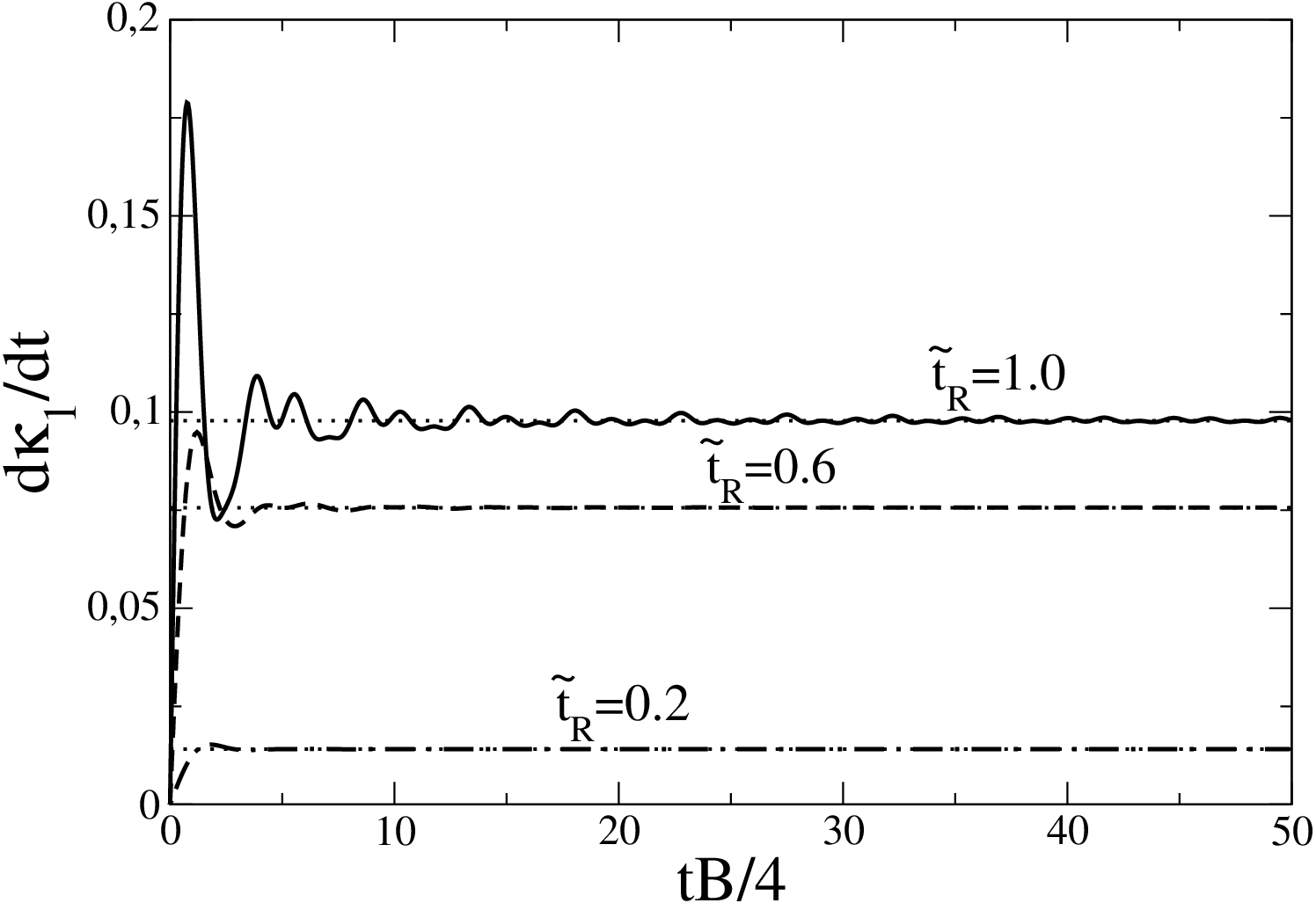}
\caption {Comparison of the exact and the approximate long time
result for $d\kappa_1/dt$ given in Eq. (\ref{kappa1p})
 for a system with  $M_L=M_R=200$, $N_{0,L}=120, N_{0,R}=100 $
and three different values of 
 $\tilde t_R$. For the smaller values of $\tilde t_R$ 
the convergence to the approximate constant result (dotted curves)
is faster. }
\label{Levitov05}
\end{center}
\vspace{0.0cm}
\end{figure}

In Fig. \ref{Levitov05} we show how the long time result for $\dot \kappa_1$
 is approached at finite times.
 From the results in Fig. \ref{Levitov04} one expects also an
 oscillatory approach of the exact numerical 
result to the Landauer-B\"uttiker formula. This is
confirmed in Fig.  \ref{Levitov05}  for
 different values of the hopping parameter between the two
subsystems.

\vspace{0.5cm}    

 For perfect transmission the long time
approximation for the first cumulant is given by
\begin{equation}
\kappa_1=\frac{t}{2\pi}\Delta \mu
\end{equation}
with $\Delta \mu\equiv \mu_L-\mu_R$. Neglecting the transition region
in Fig. 1 this is realized by $N_t$ eigenvalues $1$ of
$\bar P_{R,\rm diff}$ which implies
\begin{equation}
\label{Nt}
N_t\approx \frac{t}{2\pi}\Delta \mu~.
\end{equation}
The numerical evidence that the number of eigenvalues  $p_{m,\rm diff}(t)$
which are different from zero is in general
approximately given by $N_t$ yields for
integer  $N_t$
\begin{equation}
\kappa_1=\sum_{m=1}^{N_{\rm diff}} p_{m,\rm diff}(t)
  \approx \sum_{m=1}^{N_{t}}  p_{m,\rm diff}(t)~.
\end{equation}
This can be compared to the approximate evaluation of
Eq. (\ref{LB})
using the trapezoidal rule
for $N_t$ subintervals. Together with Eq. (\ref{Nt}) one obtains
  \begin{equation}
\label{Tm}
\kappa_1\approx 
  \sum_{m=1}^{N_{t}} T(\mu_R+\frac{m-1/2}{N_t}\Delta \mu)\equiv
 \sum_{m=1}^{N_{t}}T_m(t) ~.
\end{equation}
It is therefore tempting to conclude
  \begin{equation}
\label{LeLeguess}
 p_{m,\rm diff}(t)\approx T(\mu_R+\frac{m-1/2}{N_t}\Delta \mu)
\end{equation}
for $m=1,...,N_t$ and zero otherwise.
With Eq. (\ref{Nt}) the argument of $T$ in this equation
can also be written as $\mu_R+2\pi(m-1/2)/t$.
To really justify the guess in Eq. (\ref{LeLeguess}) one has to
examine also the higher order cumulants or compare with the exact
numerical results.

For $T=0$ the explicit form of the second order cumulant is
\begin{eqnarray}
\label{kappa2}
\kappa_2&=&
\sum_{\mu,a}\sum_{\alpha,a'}\langle k_\mu^a|
P_R(t)|k_\alpha^{a'} \rangle \langle k_\alpha^{a'}|
P_R(t)|k_\mu^a\rangle \nonumber \\
&\approx&
 \sum_{k_\mu,a}\sum_{k_\alpha,a'}
\frac{\pi^2}{(N_a+1)(N_{a'}+1)}
|\langle k_\alpha,a'+|j_{0\to 1}|k_\mu,a+\rangle
|^2 \nonumber \\
&*&\left (\frac{\sin{[(\epsilon_{k_\mu}-\epsilon_{k_\alpha})t/2]}}
{(\epsilon_{k_\mu}-\epsilon_{k_\alpha})/2}\right )^2~,
\end{eqnarray}
where $\mu$ labels occupied and $\alpha$ unoccupied one particle
states.

\begin{figure}[hbt]
\begin{center}
\vspace{-0.0cm}
\leavevmode
\epsfxsize7.5cm
\epsffile{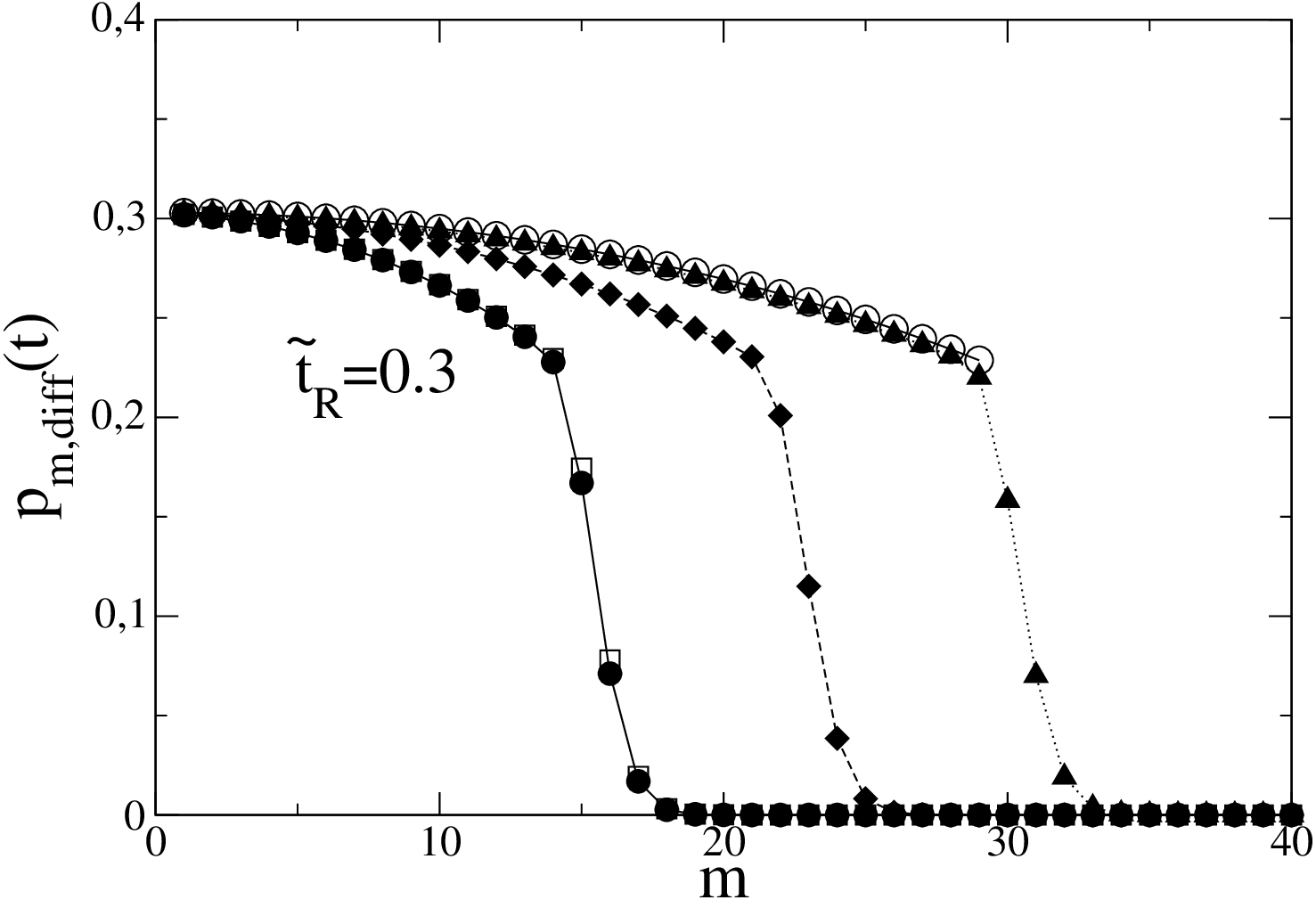}
\caption {Eigenvalues   $p_{R,\rm diff}(t)$
 for a system with  $M_L=M_R=200$, $N_{0,L}=140, N_{0,R}=100 $
and $\tilde t_R=0.3$ at $\hat t=80$ (filled circles ),
$\hat t=120$ (filled diamonds ), and
$\hat t=160$ (filled triangles ).
 For  $\hat t=80$ the eigenvalues   $p^s_{R,\rm diff}(t)$
are shown as open squares and for  $\hat t=160$ the $T_m$
defined in Eq. (\ref{Tm}) as open circles. }
\label{Levitov06}
\end{center}
\vspace{0.0cm}
\end{figure}

\noindent If one performs the thermodynamic limit and replaces the last factor
on the rhs of  Eq. (\ref{kappa2})
by $2\pi t \delta(\epsilon_{k_\mu}-\epsilon_{k_\alpha}  )$ and
 uses Eq.(\ref{current3})
one obtains Lesovik's shot noise result \cite{Leso} 
\begin{equation}
\label{kappa2zero}
\kappa_2 \approx
\frac{t}{2\pi}\int_{\mu_R}^{\mu_L} T(\epsilon)(1-T(\epsilon))d\epsilon
\end{equation}
which is the result for $\kappa_2$ which holds for the zero temperature
Levitov-Lesovik distribution in Eq. (\ref{LeLezero}). It is consistent 
with the guess in Eq. (\ref{LeLeguess}). To further test this
approximation for the eigenvalues $p_{m,\rm diff}(t)$ we show a
comparison with exact results in Fig. \ref{Levitov06}. Note that the
aproximation  $p_{m,\rm diff}(t)\approx T_m(t)$
 fails to to describe the transition region to the zero
eigenvalues. This is in contrast to the eigenvalues $p^s_{m\rm diff}(t)$
using the scattering approximation Eq. (\ref{appr2})
which at $\hat t=80$ show an overall good agreement with the exact 
$p_{m,\rm diff}(t)$.

\vspace{0.5cm}

For the special case of perfect transmission $T(\epsilon) \equiv 1 $
the approximate result in Eq. (\ref{kappa2zero})
 yields zero width for the probability
distribution $w_R(t,Q)$ in contrast to the exact
numerical result shown in Fig. \ref{Levitov01}.

In order to obtain a non zero result for $\kappa_2$ for the perfect
transmission case at $T=0$ the integrations have to be performed more
carefully. In order to keep the calculation as simple as possible we
use the approximation to only treat the electrons in the energy
window $[\mu_R,\mu_L]$ which coresponds to $N_{0,R}=0$. In addition we
work in the wide band limit
and assume an energy independent transmission probability $\bar T$,
 i.e. approximate the
current matrix element  $\langle k,L+|j_{0\to 1}|
k',L+\rangle  $ by $\bar T v_F/(2\pi)$ (see Eq. (\ref{appr3})).
 This yields for the trace of
$\bar P^2_R(t)$ entering the expression for $\kappa_2$
 in Eq. (\ref{cumulantszero})
\begin{eqnarray}
{\rm tr}\bar P^2_R(t)&=&\left (\frac{\bar T}{2\pi}\right )^2\int_0^{\Delta
  \mu}d\epsilon \int_0^{\Delta \mu}d\epsilon'\left (\frac{\sin\left
  [(\epsilon-\epsilon')t/2\right]}{(\epsilon-\epsilon')/2}\right )^2
  \nonumber \\
&=&
\frac{{\bar T}^2}{2\pi^2}\int_0^{\Delta \mu}
 d\tilde \epsilon (\Delta \mu-\tilde \epsilon)
\left ( \frac{\sin (\tilde \epsilon t/2)}
{\tilde \epsilon/2}\right )^2 \nonumber \\
&\to &
\bar T {\rm tr}\bar P_R(t)-\frac{2{\bar T}^2}{\pi^2}\int_0^{t\Delta
  \mu/2}  \frac{\sin^2u}{u} du~.
\end{eqnarray}
The remaining integral is logarithmically divergent
 in the long time limit $t\gg 1/B$ and one obtains
\begin{eqnarray}
\label{Log}
\kappa_2 \approx
\frac{t\Delta \mu}{2\pi}\bar T(1-\bar T)
+ \frac{{\bar T}^2}{\pi^2}\ln (t\Delta \mu/2) ~.
\end{eqnarray}
This agrees with the corresponding result from the approximate
solution of the matrix Riemann-Hilbert problem in Ref. \onlinecite{MA}.
The logarithmic correction term is only important for the
case of (nearly) perfect transmission. As already mentioned
for the numerical result using the exact $p_m(t)$ the cumulants
$\kappa_i$ for $i\ge 3$ vanish (apart from small transients)
for perfect transmission.
Therefore the values of $w_R(Q,t)$ lie on a Gaussian of width $\sim
 [\ln (t\Delta\mu/2)]^{1/2} $ to an excellent approximation. This
is in agreement with 
the general result for the subleading correction
for the special case considered in Ref. \onlinecite{MA}.

\subsubsection{Finite temperature results}

As the finite temperature result for $g_R(t,\lambda)$  in
Eq. (\ref{KlichLL}) contains two different functions of $\lambda$ the 
calculation of the cumulants is more tedious than for $T=0$.  
\begin{eqnarray}
\kappa_1(t)&=& {\rm tr}\left [\delta P_R(t)\bar n_0\right ] \\
\kappa_2(t)&=& {\rm tr}\left [ \delta P_R(t)\delta\bar n
- \delta P_R(t)\bar n_0 \delta P_R(t)\bar n_0\right ] \nonumber \\
&=&  {\rm tr}\left [\delta P_R(t)(\hat 1-\bar n_0)\delta P_R(t)\bar
  n_0 \right ]
\end{eqnarray}
with $\delta \bar n\equiv \bar n_{0,L}-\bar n_{0,R}$. In the second
equality for $\kappa_2$ we have used the fact that $P_R(t)=P_R+\delta P_R(t) $ is a
projection operator which reads for $\delta P_R(t)$
\begin{eqnarray}
\delta P_R(t)=P_R \delta P_R(t)+\delta P_R(t)P_R+\delta P^2_R(t)~.
\end{eqnarray}
The long time result for   $\kappa_1$ is just the Landauer-B\"uttiker
formula
integrated over time
\begin{eqnarray}
\kappa_1 &=&\sum_{i,a}\langle k_i^a|\delta P_R(t)|k_i^a\rangle
f_{k_i}^a\\
&\approx&\frac{t}{2\pi}\int T(\epsilon) \left
(f_L(\epsilon)-f_R(\epsilon)\right )d\epsilon \nonumber ~.
\end{eqnarray}
The second form for $\kappa_2$ shows explicitely its positivity
\begin{eqnarray}
&\kappa_2 &=\sum_{i,a}\sum_{j,a'}|\langle k_i^a|\delta
P_R(t)|k_j^{a'}\rangle|^2
f_{k_i}^a(1-f^{a'}_{k_j})  \\
&\approx& \frac{t}{2\pi}\int d\epsilon \{ T^2(\epsilon)\left[
  f_L(\epsilon)(1-f_L(\epsilon))+
  f_R(\epsilon)(1-f_R(\epsilon))\right ] \nonumber \\
&+& T(\epsilon)(1-T(\epsilon))\left [
  f_L(\epsilon)(1-f_R(\epsilon))+
  f_R(\epsilon)(1-f_L(\epsilon))\right ]\} \nonumber
\end{eqnarray}
The energy integral is the contribution linear in time and is obtained
by the generalization of the steps to derive Eq. (\ref{kappa2zero}).
At finite temperatures there is a contribution linear in $t$ even for
the perfect transmission case, which masks the logarithmic 
contribution in Eq. (\ref{Log}) for long enough times.

 We show in the next section
that the Levitov-Lesovik approximation  Eq. (\ref{LLformula})  amounts
 to include the
linear in $t$ contributions in all orders.   

\section{Derivation of the Levitov-Lesovik formula}

\subsection{The one-dimensional model}

After we have shown how the use of scattering states simplifies the
theoretical description in the long time limit we can use
Eq. (\ref{KlichLL}) to discuss which approximations are involved 
to derive the Levitov-Lesovik formula Eq. (\ref{LLformula}).
If we define the operator $b$ in the single particle Hilbert space
\begin{equation}
b(t,\lambda)\equiv \delta P_R(t)[\bar a_L(\lambda)- \bar a_R(\lambda)]
\end{equation}
the exact result for the characteristic function
reads
\begin{eqnarray}
\label{Spurb1}
g_R(t,\lambda)=e^{{\rm tr}\ln (1+b)}
=\exp\left \{ \sum_{m=1}^\infty \frac{(-1)^{m+1}}{m}{\rm tr}~b^m\right \}
\end{eqnarray}
In the following we simplify the exact expression for
${\rm tr}~b^m $ in two steps.

\noindent In the first step we introduce the
scattering states using Eq. (\ref{appr2}). In the second step
after taking the thermodynamic limit  the
product of the factors $(e^{i\epsilon_{j,l} t}-1)/\epsilon_{j,l}$
with $\epsilon_{j,l}=\epsilon_j-\epsilon_l$ 
is used to obtain a product of ``energy conserving'' delta functions
\begin{eqnarray} 
\label{deltafunctions}
&&\frac{e^{i\epsilon_{i_1,i_2}t}-1}{i\epsilon_{i_1,i_2} }
\frac{e^{i\epsilon_{i_2,i_3}t}-1}{i\epsilon_{i_2,i_3} }\dots
\frac{e^{i\epsilon_{i_m,i_1}t}-1}{i\epsilon_{i_m,i_1} } \\
&&=\frac{\sin{(\epsilon_{i_1,i_2}t/2)}}{\epsilon_{i_1,i_2}/2 }
\frac{\sin{(\epsilon_{i_2,i_3}t/2)}}{\epsilon_{i_2,i_3}/2 } \dots
\frac{\sin{(\epsilon_{i_m,i_1}t/2)}}{\epsilon_{i_m,i_1}/2 }\nonumber \\
&& \to~(2\pi)^{m-1} t\delta(\epsilon_{i_1}-\epsilon_{i_2} )
\delta(\epsilon_{i_2}-\epsilon_{i_3} )\dots
 \delta(\epsilon_{i_m}-\epsilon_{i_1} ) \nonumber
\end{eqnarray}
The $k$- integrations in the thermodynamic limit can be converted to 
energy integrations. Because of the product of delta functions
only one energy integration remains. The summations
over the quantum numbers $L,R$ still have to be performed.
 They can be expressed as a trace in a
$2$-dimensional space spanned by $a=L,R$
\begin{eqnarray}
\label{Spurb2} 
{\rm tr}~b^m \to \frac{t}{2\pi}\int 
d\epsilon~ {\rm tr}_{(2)}[c_{(2)}(\epsilon,\lambda)]^m
\end{eqnarray}
with
\begin{eqnarray} 
\!\!\!&&c_{(2)}(\epsilon,\lambda) =\\ 
\!\!\!&& \frac{2\pi}{ v(\epsilon) }\left (
\begin{matrix}
\langle k,L+|j| k,L+\rangle a_L
~~-\langle k,L+|j| k,R+\rangle a_R\\
\langle k,R+|j| k,L+\rangle a_L
~~-\langle k,R+|j| k,R+\rangle a_R
\end{matrix} 
\right ) \nonumber \\
\!\!\!&&=
\left(
\begin{matrix}
~|t_k|^2a_L(\epsilon,\lambda)~~-t_k^*\tilde r_ka_R(\epsilon,\lambda)\\
t_k \tilde r_k^*a_L(\epsilon,\lambda)~~~~ |\tilde t_k|^2a_R(\epsilon,\lambda)
\end{matrix} \nonumber
\right )~.
\end{eqnarray}
Here $k=k(\epsilon)$,
$a_{L(R)}(\epsilon,\lambda)  \equiv f_{L(R)}(\epsilon)(e^{\pm i\lambda}-1)$
 and we have dropped the index $0\to 1$ of the 
current operator.

 From Eqs. (\ref{Spurb1}) and (\ref{Spurb2}) one obtains
\begin{eqnarray} 
\label{twobytwo}
{\rm tr}[\ln(1+b)]&\to& \frac{t}{2\pi} \int d\epsilon \sum_{m=1}^\infty
\frac{(-1)^{m+1}}{m}{\rm tr_{(2)}}[c_{(2)}(\epsilon,\lambda)]^m
\nonumber \\
&=& \frac{t}{2\pi} \int d\epsilon~ {\rm tr_{(2)}}[\ln(
\hat 1_2+c_{(2)}(\epsilon,\lambda)) ]\nonumber \\
&=& \frac{t}{2\pi} \int d\epsilon \ln[{\rm det_{(2)}}(\hat 1_2+c_{(2)}
(\epsilon,\lambda)  )].
\end{eqnarray}
The $2\times2$ determinant can easily be calculated. Using $|\tilde
r_k|^2=1-T(\epsilon)$ and $a_La_R=-(f_Ra_L+f_La_R)$ one obtains
\begin{eqnarray} 
{\rm det_{(2)}}(\hat 1_2+c_{(2)})&=&1+ T(\epsilon)[(e^{i\lambda}-1)
f_L(\epsilon)(1-f_R(\epsilon)) \nonumber \\
 \!\!\!&+&   (e^{-i\lambda}-1)f_R(\epsilon)(1-f_L(\epsilon)) ] ~.
\end{eqnarray}
If one inserts this into Eq. (\ref{twobytwo}) this completes the
derivation of the Levitov-Lesovik formula. Note that in the derivation
it was not necessary to invoke the wide band limit.

If one is interested in logarithmic corrections in time  to $\ln
g_R(t,\lambda)$ the approximation in Eq. (\ref{deltafunctions})
should not be made for carrying out the energy integrations (see
the discussion of $\kappa_2$ leading to Eq. (\ref{Log}) in section IV b).

\subsection{Multi-lead geometries}
\begin{figure}[hbt]
\begin{center}
\vspace{0.5cm}
\leavevmode
\epsfxsize7.5cm
\epsffile{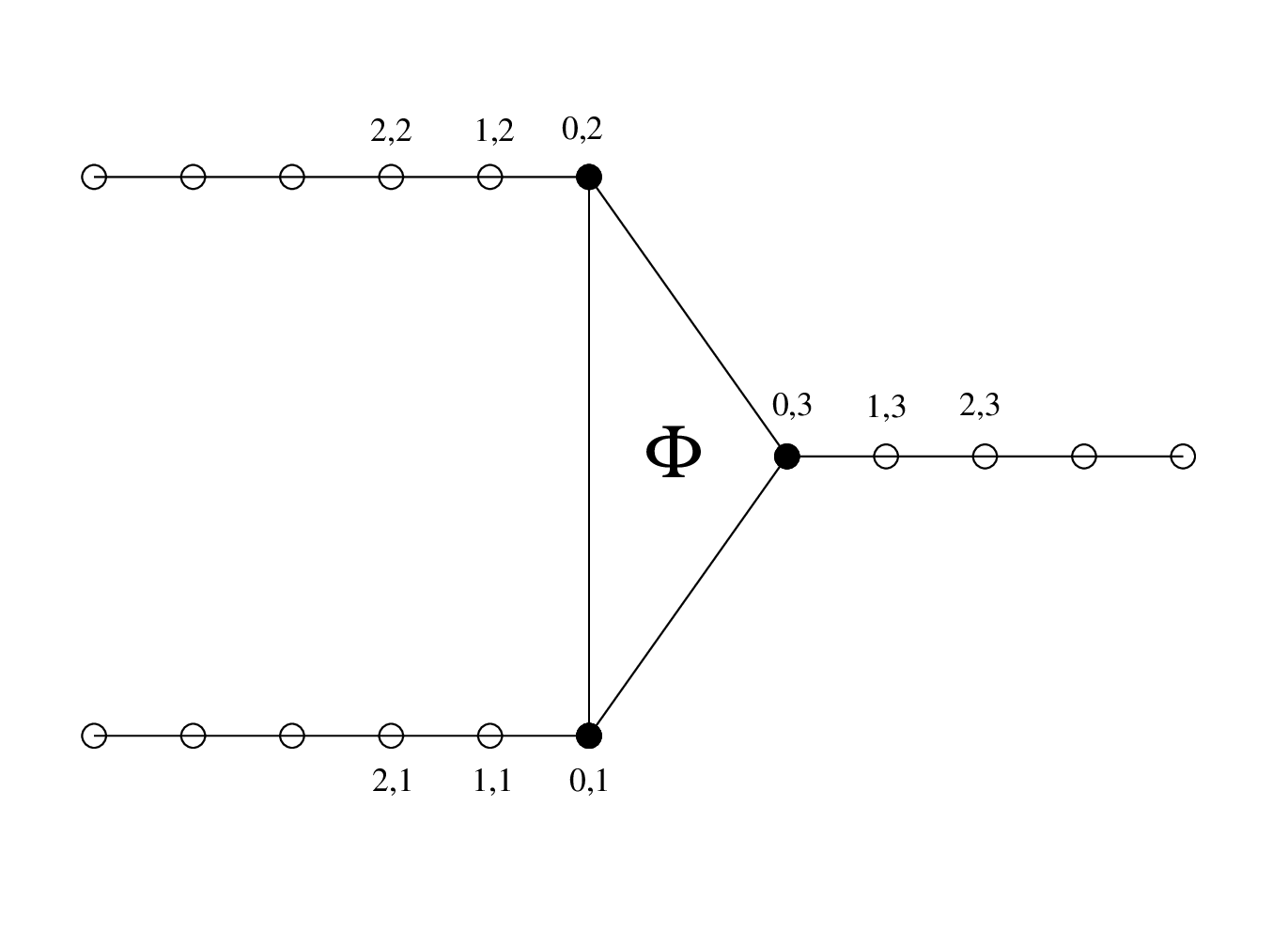}
\caption {Junction of $M=3$ quantum wires connected to a ``ring''
indicated by the filled circles, which is
pierced by a magnetic flux. In the initial state the hopping matrix
elements between the ring sites and the adjacent wire sites vanish.    }
\label{Levitov07}
\end{center}
\vspace{0.0cm}
\end{figure}

The FCS in multiterminal circuits was addressed
 by various authors.\cite{NB,BBB}
For noninteracting electrons 
the Levitov-Lesovik formula for the quantum wire discussed in the previous
subsection can easily be generalized to such more complex geometries.
As an example we show in Fig. \ref{Levitov07} a Y-junction
pierced by a magnetic flux. 
The zero temperature limit of a Y-junction was already discussed in
ref. 3. 
The proof given below works 
 for the general case of $M$ legs at finite temperatures.  

\vspace{0.1cm}

We denote the states on the central ``ring'' as $|0,a\rangle$ with 
$a=1,...M$ and the states in the legs in the site representation
as $|m,a\rangle$ with $m=1,2,...\infty$. The connection of the legs 
with the ring generalizes Eq. (\ref{connection}) to 
\begin{equation} 
\label{interaction}
v=-\sum_{a=1}^M \tau_a(|0,a\rangle \langle 1,a| +H.c.) ,
\end{equation}
where the $ \tau_a$ can be assumed to be real.
In the legs we assume nearest neighbor hopping $\tau=1$.
At time $t=0$ the subsystems with different chemical potentials
(and temperatures) are connected via the hopping term $v$ in
Eq. (\ref{interaction}).
Instead of the particle number in the right part of the system we 
now monitor the particle number in leg $M$. The relevant current operator
is then given by
\begin{equation} 
j_{0\to 1,M}=i\tau_M (|1,M\rangle \langle 0,M|- |0,M\rangle \langle 1,M| )~.
\end{equation}
The scattering states take the form
\begin{eqnarray}
\label{StreuL}
\langle m, a'|k,a+\rangle
&=&\frac{1}{\sqrt{2\pi}}(e^{-ikm}+r_{a,k}e^{ikm})~,~ a'= a \nonumber \\
&=&\frac{1}{\sqrt{2\pi}}~t^{a'\leftarrow a}_ke^{ikm}~, ~a'\ne a~,
\end{eqnarray} 
where $m\ge 1$. When the ring encloses a magnetic flux 
 $|t^{a'\leftarrow a}_k|$
 generically differs from  $|t^{a\leftarrow a'}_k|$  . \cite{Xavier}.

In order to calculate the current matrix elements we have to generalize
Eq. (\ref{zerosite}) to
\begin{eqnarray}
\tau_M\langle 0,M|k,a+\rangle &=&\frac{1}{\sqrt{2\pi}}~t^{M\leftarrow
  a}_k~, ~~ a\ne M\\
\tau_M\langle 0,M|k,M+\rangle &=&\frac{1}{\sqrt{2\pi}}(1+r_{M,k})~. \nonumber
\end{eqnarray}
Having defined the multi-lead model we can now generalize our 
description of the FCS in section III . The operator $\bar n_0$
now takes the form
\begin{equation}
\bar n_0=\bar n_{0,\rm ring}+\sum_{a=1}^M \bar n_{0,a}\equiv \bar n_0'
+ \bar n_{0,M}
\end{equation}
and  the operator $b$ in  the  
characteristic function $g_M(t,\lambda)={\rm det}(\hat 1 +b(t,\lambda))$
 is given by
\begin{eqnarray}
 b(t,\lambda)   =                       \delta P_M(t) \left
[\bar n_0'(e^{i\lambda}-1)- \bar n_{0,M} (e^{-i\lambda}-1)\right ]~.
\end{eqnarray}
As in the preceeding subsection we evaluate the determinant in the 
long time limit by calculating ${\rm tr}~b^m$ to all orders. 
As we are only interested in the linear in $t$ terms
we can neglect the ring state contribution to the trace.  We then  obtain
as the generalization of Eq. (\ref{Spurb2})  
\begin{eqnarray}
\label{SpurbM} 
{\rm tr}~b^m \to \frac{t}{2\pi}\int 
d\epsilon~ {\rm tr}_{(M)}[c_{(M)}(\epsilon,\lambda)]^m
\end{eqnarray}
The current matrix elements entering the matrix $c_{(M)}(\epsilon,\lambda) $ 
are given by
\begin{eqnarray}
\langle k,\tilde a+|j_{0\to 1,M}| k,\tilde a'+\rangle
&=& \frac{v(k)}{2\pi} (t^{M\leftarrow
 \tilde  a}_k)^* t^{M\leftarrow \tilde  a'}_k \nonumber \\
\langle k,\tilde a+|j_{0\to 1,M}| k,M+\rangle
&=& \frac{v(k)}{2\pi} (t^{M\leftarrow
 \tilde  a}_k)^* r_{M,k}  \\
\langle k,M+|j_{0\to 1,M}| k,M+\rangle
&=& -\frac{v(k)}{2\pi}(1-|r_{M,k}|^2) \nonumber~.
\end{eqnarray}
Here $\tilde a,\tilde a'\ne M$.
For the Y-junction with $M=3$ the matrix $\hat 1_3+c_{(3)}(\epsilon,\lambda) $
is given by 
\begin{eqnarray}
\hat 1_3+c_{(3)}=\left (
\begin{matrix}
1+t_1^*a_1~~~~t_1^*a_2~~~~~~-t_1^*a_3\\
~~~t_2^*a_1~~~~1+ t_2^*a_2~~~-t_2^*a_3 \\
~~r^*a_1~~~~~r^*a_2~~~~~~~1+D
\end{matrix} 
\right )~.
\end{eqnarray}
Here we have introduced the abbreviations $t^{M\leftarrow a}_k\to t_a$
and $r_{M,k} \to r$ and defined
 $a_{1(2)}=t_{1(2)}f_{1(2)}
(e^{i\lambda}-1), a_3=rf_3(e^{-i\lambda}-1)$
and $D=(1-|r|^2)f_3 (e^{-i\lambda}-1)$. As for general values of $M$
the determinant of this matrix is best calculated by expanding in the
last row. This yields
\begin{equation}
{\rm det}(\hat 1_3+c_{(3)})=
(1+D)(1+t^*_1a_1+t^*_2a_2)+r^*a_3(t^*_1a_1+t^*_2a_2).
\end{equation}
For arbitrary values of $M$ the sum over the $t^*_ia_i$ extends to
$M-1$.
Therefore   the generalization to arbitrary
values of $M$ reads with $d_{\pm}\equiv e^{\pm i\lambda}-1$
and using $d_+d_-=-(d_++d_-)$
\begin{eqnarray}
\label{detM}
{\rm det}(\hat 1_M+c_{(M)})&=&
1+d_+(1-f_M) \sum_{a=1}^{M-1} |t^{M\leftarrow a}_k|^2 f_a  \\
&+&  d_-f_M \left (1-|r_{M,k}|^2-\sum_{a=1}^{M-1}  |t^{M\leftarrow a}_k|^2
 f_a\right ). \nonumber
\end{eqnarray}
The conservation of probability implies
\begin{equation}
1-|r_{M,k}|^2= \sum_{a=1}^{M-1} |t^{a\leftarrow M}_k|^2
\end{equation}
As generally $ |t^{M\leftarrow a}_k|^2\ne |t^{a\leftarrow M}_k|^2$
it requires an additional argument to bring Eq. (\ref{detM})
into its final form. It is shown in the appendix B that
\begin{equation}
\label{relation}
\sum_{a=1}^{M-1} |t^{a\leftarrow M}_k|^2=
\sum_{a=1}^{M-1}|t^{M\leftarrow a}_k|^2  
\end{equation}
also holds in the presence of a magnetic flux. Using
Eq. (\ref{twobytwo}) for arbitrary 
values of $M$ we obtain the generalized Levitov-Lesovik formula .
With
$T_a(\epsilon)\equiv |t^{M\leftarrow a}_k|^2$ it reads 
\begin{equation} 
\ln{g_M(t,\lambda}) \to \frac{t}{2\pi}\int d\epsilon
\ln{\left [1+A(\epsilon,\lambda)\right]}
\end{equation}
with
\begin{eqnarray}
A(\epsilon,\lambda)
&=&\sum_{a=1}^{M-1}T_a(\epsilon)\left
[(e^{i\lambda}-1)f_a(\epsilon)(1-f_M(\epsilon))  \right.\nonumber \\
&+&(e^{-i\lambda}-1)f_M(\epsilon)(1-f_a(\epsilon)) \left. \right ]
\end{eqnarray}
Again it was not necessary to invoke the wide band limit.
\section{Summary}

The aim of this paper was to clarify the validity of the Levitov-Lesovik
 formula for the FCS of noninteracting fermions and to present
a simple derivation which clearly shows the aproximations involved. 
As a starting point we used the formula for the characteristic
function in Eq. (\ref{Kli}) first derived by Klich. \cite{Klich}
No measuring device is used in this description. 
 As Klich noted his formal result
 should be viewed as the general
expression for the counting statistics of noninteracting fermions,
at any given time and without any approximation.
 Our alternative derivation of
Klich's formula for 
finite temperatures presented in Appendix A uses Wick's theorem.

The first step in the actual calculation of the characteristic
function is to obtain an explicit result for the time dependent
projection 
operator $P_R(t)$. This was done exactly
for a microscopic lattice model
 using a numerical approach
and analytically in the long time limit using scattering states.
It was shown by a comparison of the two aproaches that the analytical
expression in Eq. (\ref{appr2}) works very well after transients
with a time scale $\sim 1/B$ have died out.    
At zero temperature
 the eigenvalues $p_m(t)$ of the operator $\bar P_R(t)$ directly
determine the probability distribution $w(Q,t)$. The fact that the Levitov-Lesovik
formula provides only an approximate description of the FCS 
was studied in detail using low order cumulants. A
comparison to results obtained by different methods \cite{MA} was
given.

In section VI we presented a straightforward derivation of the Levitov-Lesovik
formula summing all linear in $t$ contributions
to $\ln g_R(t,\lambda)$ using
Eqs. (\ref{Kli}) and  (\ref{appr2}).
An important step in the derivation was to first bring 
Eq. (\ref{Kli}) into the form Eq. (\ref{KlichLL})
 introducing the operator $\delta P_R(t)$.
In the derivation no assumptions about the band width an the energy
dependence of the transmission probability were necessary.
The derivation applied to the simple ``experimental'' setup studied
in this paper shows that the Levitov-Lesovik formula provides an excellent
approximation in the long time limit, except in the case of (near)
perfect transmission at very low temperatures.
 An extension of this new
derivation to multi-lead circuits with possible broken
time reversal symmetry was also given.

\section{Acknowledgements}

The author is grateful to A. Gogolin for discussions on
how to best derive the Levitov-Lesovik formula.  
 
\begin{appendix}

\section { Klich's formula using Wick's theorem}

In this appendix we discuss the expection value of an exponential 
of a selfadjoint one-particle operator $A$
 \begin{equation} 
A=\sum_{i,j}\langle i|\hat a|j\rangle c_i^\dagger c_j=\sum_\lambda a_\lambda
 c_\lambda^\dagger c_\lambda
\end{equation}
where the $c_i^{(\dagger)}$ (as well as the 
$c_\lambda^{(\dagger)}$ ) obey the usual anticommutation relations
$\{c_i,c_j^\dagger \}=\delta_{ij}, \{c_i,c_j \}=0$. In order to avoid 
mathematical subtleties we assume the dimension of the one-particle
Hilbert space to be finite $(M)$. The creation operators for the 
two different one-particle basis states obey the usual relation
 \begin{equation} 
 c_\lambda^\dagger=\sum_i c_i^\dagger \langle i |\lambda\rangle ~.
\end{equation}

We consider expectation values with a statistical operator
in Fock space
\begin{equation} 
\rho_0=e^{-\bar H_0}/{\rm Tr}e^{-\bar H_0}\equiv e^{-\bar H_0}/\bar Z_0~, 
\end{equation}
with
\begin{equation} 
\bar H_0=\sum_i\bar \epsilon_i c_i^\dagger c_i~.
\end{equation}
In the section II we use
$\bar H_0=\beta_L(H_{0,L}-\mu_L{\cal N}_L)
+\beta_R(H_{0,R}-\mu_R{\cal N}_R)  $, where the $H_{0,a} $ are
one-particle operators and the ${\cal N}_a $ are the particle number
operators of the left and right part of the system. 
The grand canonical partion function $\bar Z_0$ is given by
\begin{equation} 
\bar Z_0=\prod _i(1+e^{-\bar \epsilon_i})={\rm det}(1+e^{-\bar h_0})~.
\end{equation}
 With the operator in the one particle Hilbert space
\begin{equation} 
\bar n_0\equiv (e^{\bar h_0}+1)^{-1}=\sum_i|i\rangle \frac{1}{e^{\bar
    \epsilon_i}+1}\langle i|
\end{equation}
the expectation values of bilinear operators in the
$\lambda$-representation
 can be written as
\begin{equation} 
\label{bi}
\langle c^\dagger_\lambda c_\mu \rangle=
\sum_{i,j}\langle \mu|j\rangle \langle c^\dagger_i c_j \rangle
\langle i|\lambda \rangle=
 \langle \mu|\bar
n_0|\lambda\rangle~.
\end{equation}
We now consider the expectation value of $e^{\alpha A}$, where $\alpha$
is an arbitrary complex number
\begin{equation} 
\langle e^{\alpha A}\rangle= \langle e^{\alpha \sum_\lambda a_\lambda
 c_\lambda^\dagger c_\lambda }\rangle= \prod_\lambda\left \langle
\left[1+(e^{\alpha a_\lambda }-1)c^\dagger_\lambda c_\lambda  
 \right ]\right \rangle~.
\end{equation}
The product runs over the $M$ different eigenstates labeled by
$\lambda$. If one multiplies out the product one has to caclulate
expectation values of products of different $\hat n_\lambda=
c_{\lambda}^\dagger c_{\lambda}  $. They are given by Wick's theorem
\cite{FW} as a determinant of the bilinear expectation values in
Eq. (\ref{bi})
\begin{eqnarray} 
\label{Wick1}
\langle c_{\lambda_1}^\dagger c_{\lambda_1}....c_{\lambda_m}^\dagger
c_{\lambda_m} \rangle=
\left | 
\begin{matrix}
\langle c_{\lambda_1}^\dagger c_{\lambda_1}\rangle &\dots& \langle
 c_{\lambda_1}^\dagger
c_{\lambda_m}\rangle \\
\vdots &\dots &\vdots \\
\langle c_{\lambda_m}^\dagger c_{\lambda_1}\rangle &\dots& \langle 
c_{\lambda_m}^\dagger
c_{\lambda_m } \rangle
\end{matrix} 
\right |
 \end{eqnarray}  
This expectation value is multiplied by $\prod_{i=1}^m
 (e^{\alpha a_{\lambda_i} }-1)   $. 
These factors can be incorporated into the matrix elements on the rhs
of Eq. (\ref {Wick1}) using Eq.  (\ref {bi})
\begin{equation}
 (e^{\alpha a_{\lambda_j} }-1)\langle c^\dagger_{\lambda_i}
 c_{\lambda_j} \rangle=
\langle \lambda_j|(e^{\alpha\hat a}-1)\bar n_0
|\lambda_i\rangle~.
\end{equation}
A compact expression for $\langle e^{\alpha A}\rangle$
 is finally obtained by comparison
with the formula for the 
determinant of $\hat 1 +\hat b$, where $\hat b$ is an
arbitrary linear operator in the $M$-dimensional Hilbert space
\begin{eqnarray}
{\rm det}(\hat 1 +\hat b )&=&
1+{\rm tr} ~\hat b+\sum_{i<j}{\rm det}^{(2)}b^{(2)}
+\sum_{i<j<k}{\rm det}^{(3)}b^{(3)} \nonumber \\
&+& .....+{\rm det}(\hat b )~,
\end{eqnarray}
where e.g. ${\rm det}^{(3)}b^{(3)} $  denotes a $3\times 3$
subdeterminant of $\hat b$ with the indices given by the summation
variables. The comparison with the calculation of  $\langle e^{\alpha A}\rangle$
yields
\begin{equation}
\label{expec}
\langle e^{\alpha A}\rangle={\rm det}\left [\hat 1 +(e^{\alpha \hat
    a}-1)\bar n_0 \right ]~.
\end{equation}
This result also follows from Klich's trace formula\cite{Klich}
\begin{equation}
\label{Klichf}
{\rm Tr} (e^Ae^B)={\rm det}(\hat 1 +e^{\hat a}e^{\hat b})~,
\end{equation}
where $A$ and $B$ are arbitrary one particle operators in Fock space
and $\hat a$ and $\hat b$ are the corresponding operators in the
Hilbert space of a single particle.
The proof of this formula is nontrivial when the operators $A$ and $B$
do not commute.
 As Klich's
elegant proof \cite{Klich} involves
a result not widely known and it is not availble in a 
journal publication, we presented the alternative proof of Eq. (\ref{expec}) 
using Wick's theorem to make this paper selfcontained.
In section II we use a slight generalization of
Eq. (\ref{expec}).

Let $C$ be a one particle operator in Fock space which commutes
with $\bar H_0$. With $\bar H_0^{'} \equiv \bar H_0-\gamma C $ we obtain
\begin{eqnarray}
\label{expecgeneral}
\langle e^{\alpha A}e^{\gamma  C}\rangle&=&
\frac{{\rm Tr}  e^{\alpha A} e^{-\bar H_0^{'}}}
{{\rm Tr}e^{-\bar H_0^{'}}  }
~\frac{{\rm Tr}e^{-\bar H_0^{'}}  }{{\rm Tr}e^{-\bar H_0}  } \nonumber \\
&=& {\rm det}\left [\hat 1- \bar n_0^{'} +e^{\alpha \hat a}\bar n_0^{'}\right ]
\frac{{\rm det}(\hat 1 +e^{-\bar h_0^{'}} ) }{{\rm det}(\hat 1 +e^{-\bar
    h_0} ) } \nonumber \\
&=&{\rm det}\left [\hat 1+( e^{\alpha \hat a}e^{\gamma \hat c}-1
  )\bar n_0\right ]
\end{eqnarray}

\section{Proof of Eq. (\ref{relation})}

The scattering states can expressed via the full resolvent
$g(z)= (z-h)^{-1}$
and the unperturbed resolvent $g_0(z)=  (z-h_0)^{-1}$ which are
related by $g=g_0+gvg_0$. From the definition in Eq. (\ref{scatstate})
one obtains for $a'\ne a$
\begin{eqnarray}
\langle m,a|k,a'+\rangle&=&\langle
m,a|g_0(\epsilon_k+i0)|1,a\rangle\tau_a\\
&\times&
\langle 0,a|g(\epsilon_k+i0)|0,a'\rangle \tau_{a'}\langle
0,a'|k,a'\rangle~.
\nonumber
\end{eqnarray}
The resolvent matrix element of the semi-infinite chain is given by
\cite{Enns} $\langle
m,a|g_0(\epsilon_k+i0)|1,a\rangle=-e^{-ikm}$. With  Eqs. 
(\ref{standing1}), (\ref{standing2}) and (\ref{StreuL})
the transmission probabilities follow as
\begin{eqnarray}
\label{transM}
|t^{a'\leftarrow a}_k|^2=4\tau_a^2\tau^2_{a'}\sin^2k|\langle
 0,a'|g(\epsilon_k+i0)|0,a\rangle|^2~,
\end{eqnarray}
Using the projection $P_r$ onto the states on the ring the full resolvent 
matrix elements in Eq. (\ref{transM}) can be written as \cite{Enns}
\begin{eqnarray}
 \langle 0,a'|g(z)|0,a\rangle = \langle 0,a'|[zP_r-h_r-\gamma(z)]^{-1}|0,a
\rangle
\end{eqnarray} 
with
\begin{eqnarray}
\label{Gamma}
\gamma(z)=g_b^0(z)\sum_{a=1}^M\tau_a^2|0,a\rangle \langle 0,a|~,
\end{eqnarray} 
where $h_r$ is the one particle Hamiltonian
on the ring and $ g_b^0(z)\ $ is the diagonal element of the resolvent
of the semi-infinite chain at the boundary. For $z=\epsilon\pm i0$ it is
given by
\begin{eqnarray}
 g_b^0(\epsilon\pm i0)=(\epsilon\mp i\sqrt{4-\epsilon^2})/2~.
\end{eqnarray} 
In order to prove the relation between the 
transmission probabilities in Eq. (\ref{relation}) we us a simple
operator identity. Let $a$ and $b$ operators such that the inverses of
$a\pm b$ exist. The one has
\begin{eqnarray}
\frac{1}{a-b}-\frac{1}{a+b}&=& \frac{1}{a-b}(a+b)\frac{1}{a+b}-
\frac{1}{a-b}(a-b)\frac{1}{a+b} \nonumber \\
&=& 2~\frac{1}{a-b}~b~\frac{1}{a+b}~.
\end{eqnarray} 
If one puts the unit operators $(a\pm b) (a\pm b)^{-1}$ on the
opposite sides one obtains by comparison the operator relation
\begin{eqnarray}
\frac{1}{a-b}~b~\frac{1}{a+b}=\frac{1}{a+b}~b~\frac{1}{a-b}
\end{eqnarray} 
which we now use 
for operators acting in the subspace of the ring states.
With $a=\epsilon \hat 1-h_r-{\rm Re} \gamma(\epsilon+i0)$ and $b={\rm Im}
\gamma(\epsilon+i0)\equiv \gamma_{\rm I}(\epsilon) $ the relation
reads using ${\rm Re} \gamma(\epsilon+i0)={\rm Re}
\gamma(\epsilon-i0)$
and ${\rm Im}\gamma(\epsilon+i0)=-{\rm Im}\gamma(\epsilon-i0)$
\begin{eqnarray}
g(\epsilon+i0)\gamma_{\rm I}(\epsilon)g(\epsilon-i0)=
g(\epsilon-i0)\gamma_{\rm I}(\epsilon)g(\epsilon+i0).
\end{eqnarray} 
We next take the expectation value in the state $|0,M\rangle$ and
use the explicit form of $\gamma(z)$ in Eq. (\ref{Gamma}). The $M=a$
term of the sum can be omitted on both sides of the equation as they
are identical. Deviding by $  g_b^0(\epsilon+ i0) $ yields
\begin{eqnarray}
\sum_{a=1}^{M-1}\tau_a^2|\langle 0,M|g(\epsilon+i0)|0,a\rangle |^2
=\sum_{a=1}^{M-1}\tau_a^2|\langle 0,a|g(\epsilon+i0)|0,M \rangle |^2 \nonumber
\end{eqnarray} 
Together with the explicit result for the transmission probalities in
Eq. (\ref{transM}) this completes the proof of the relation between the
transmission probabilities in Eq. (\ref{relation}).

\end{appendix}

\end{document}